\documentclass[sn-mathphys]{sn-jnl}

\jyear{2022}%

\theoremstyle{thmstyleone}%
\theoremstyle{thmstyletwo}%

\theoremstyle{thmstylethree}%

\newcommand{\bc}{{\bf c}}
\newcommand{\bx}{{\bf x}}
\newcommand{\bz}{{\bf z}}
\newcommand{\bbeta}{\mbox{\boldmath $\beta$}}

\newcommand{\beps}{\mbox{\boldmath $\epsilon$}}
\newcommand{\bVareps}{\mbox{\boldmath $\varepsilon$}}
\newcommand{\brho}{\mbox{\boldmath $\rho$}}
\newcommand{\btheta}{\mbox{\boldmath $\theta$}}
\newcommand{\bpsi}{\mbox{\boldmath $\psi$}}
\newcommand{\bLambda}{\mbox{\boldmath $\Lambda$}}
\newcommand{\bPsi}{\mbox{\boldmath $\Psi$}}
\newcommand{\bSigma}{\mbox{\boldmath $\Sigma$}}
\newcommand{\bOmega}{\mbox{\boldmath $\Omega$}}
\newcommand{\bC}{{\bf C}}
\newcommand{\bI}{{\bf I}}
\newcommand{\bL}{{\bf L}}
\newcommand{\bW}{{\bf W}}
\newcommand{\bZ}{{\bf Z}}
\newcommand{\zero}{{\bf 0}}

\newcommand{\pmat}{\mbox{\boldmath $\mathcal{P}$}}

\begin{document}

\title[Bayesian MMP with Sparse Response]{Bayesian Analysis of Multivariate Matched Proportions with Sparse Response}


\author*[1]{\fnm{Mark J.} \sur{Meyer}}\email{mjm556@georgetown.edu}

\author[1]{\fnm{Haobo} \sur{Cheng}} 

%
\author[2]{\fnm{Katherine Hobbs} \sur{Knutson}} 

\affil*[1]{\orgdiv{Department of Mathematics and Statistics}, \orgname{Georgetown University}, \orgaddress{\city{Washington}, \state{DC}, \postcode{20057}, \country{USA}}}

%
\affil[2]{\orgname{Duke University School of Medicine}, \orgaddress{\city{Durham}, \state{NC}, \postcode{27710}, \country{USA}}}



\abstract{Multivariate matched proportions (MMP) data appears in a variety of contexts including post-market surveillance of adverse events in pharmaceuticals, disease classification, and agreement between care providers. It consists of multiple sets of paired binary measurements taken on the same subject. While recent work proposes non-Bayesian methods to address the complexities of MMP data, the issue of sparse response, where no or very few ``yes'' responses are recorded for one or more sets, is unaddressed. The presence of sparse response sets results in underestimates of variance, loss of coverage, and lowered power in existing methods. Bayesian methods have not previously been considered for MMP data but provide a useful framework when sparse responses are present. In particular, the Bayesian probit model provides an elegant solution to the problem of variance underestimation. We examine three approaches built on that model: a na\"ive analysis with flat priors, a penalized analysis using half-Cauchy priors on the mean model variances, and a multivariate analysis with a Bayesian functional principal component analysis (FPCA) to model the latent covariance. We show that the multivariate analysis performs well on MMP data with sparse responses and outperforms existing non-Bayesian methods. In a re-analysis of data from a study of the system of care (SOC) framework for children with mental and behavioral disorders, we are able to provide a more complete picture of the relationships in the data. Our analysis provides additional insights into the functioning on the SOC that a previous univariate analysis missed.
}


\maketitle

\section{Introduction}

Multivariate matched proportions (MMP) data arises when multiple sets of paired binary measurements are taken on the same subject. An example of such data comes from a study of care coordination within a System of Care (SOC) framework \citep{Knutson2018}. An SOC informs the treatment of children with severe mental and behavioral disorders by coordinating care between six different components: mental health, primary care, the education system, child welfare, juvenile justice, and developmental disability services. However, primary care frequently provides the sole treatment for children with mental health disorders making it unclear if those children are receiving the potential benefits of the SOC. In particular, primary care may not be initiating contact with SOC components at the same rate as psychiatric care. To begin to investigate primary care's tendency to initiate coordination within an SOC, Knutson et al. \cite{Knutson2018} present the results of a retrospective chart study of youth referred by pediatricians to a child psychiatrist at an urban community health center. The authors assess coordination by examining the documented contacts in the medical record between primary care and each of the remaining components of the SOC. The study then compares those contacts to the documented contacts between the psychiatrist, or specialty care, and the SOC components made after the initial psychiatric evaluation.

We summarize the contact data for all of the SOC components in Table~\ref{t:soc} where the column headers denote the assessor, primary care or specialty care, and the column sub-headers denote the components. The original analysis of this data did not account for the multivariate nature of the data nor did it account for the sparse responses in contacts made with two components. Sparse response is most noticeable for contact made with developmental disabilities by specialty care where no contacts were made (see Table~\ref{t:soc}). The sparse response of juvenile justice requires noting that not all combinations of zeros and ones appear in the data, only $0, 0$ and $0, 1$. Alternatively, we can represent the data in a series of population averaged tables, as in Table~\ref{t:popavg}, and diagnose sparse response by noting the zero cell counts in both developmental disabilities and juvenile justice tables. Since each component has a corresponding set of paired binary measurements, two of the five sets in this data exhibit sparse response.

	\begin{table}
		\centering
			\caption{Multivariate matched proportions data from the SOC study for care-specific contact with developmental disabilities (DD), mental health (MH), juvenile justice (JJ), child welfare (CW), and education system (ED). Table values are 1 if contact was made with the agency in the column sub-header by the assessor in the column header, and 0 if no contact was made. The Pattern Count column denotes the number of times the pattern appears in the data. This table excludes patterns that do not appear in the data.}
			\label{t:soc}
		\begin{tabular}{ccccccccccccc}
			\toprule
			 \multicolumn{5}{c}{Primary Care} & &  \multicolumn{5}{c}{Specialty Care} &  & Pattern\\ 
			 \cmidrule{1-5} \cmidrule{7-11}
			 DD & MH & JJ & CW & ED &  & DD & MH & JJ & CW & ED &  & Count \\ 
			  \midrule
			 0 & 0 & 0 & 0 & 0 &  & 0 & 0 & 0 & 0 & 1 &  & 17 \\ 
			 0 & 0 & 0 & 0 & 0 &  & 0 & 0 & 0 & 0 & 0 &  & 16 \\ 
			 0 & 1 & 0 & 0 & 0 &  & 0 & 1 & 0 & 0 & 0 &  & 5 \\ 
			 0 & 0 & 0 & 0 & 1 &  & 0 & 0 & 0 & 0 & 1 &  & 5 \\ 
			 0 & 1 & 0 & 0 & 0 &  & 0 & 1 & 0 & 0 & 1 &  & 3 \\ 
			 0 & 0 & 0 & 0 & 0 &  & 0 & 1 & 0 & 0 & 0 &  & 3 \\ 
			 0 & 0 & 0 & 0 & 0 &  & 0 & 1 & 0 & 0 & 1 &  & 3 \\ 
			 0 & 1 & 0 & 1 & 0 &  & 0 & 0 & 0 & 0 & 0 &  & 2 \\ 
			 0 & 0 & 0 & 0 & 1 &  & 0 & 0 & 0 & 0 & 0 &  & 2 \\ 
			 0 & 0 & 0 & 0 & 0 &  & 0 & 0 & 0 & 1 & 1 &  & 2 \\ 
			 1 & 0 & 0 & 0 & 0 &  & 0 & 0 & 0 & 0 & 0 &  & 1 \\ 
			 0 & 1 & 0 & 0 & 0 &  & 0 & 0 & 0 & 0 & 0 &  & 1 \\ 
			 0 & 1 & 1 & 0 & 0 &  & 0 & 1 & 1 & 1 & 0 &  & 1 \\ 
			 0 & 1 & 0 & 0 & 1 &  & 0 & 1 & 0 & 1 & 1 &  & 1 \\ 
			 0 & 1 & 0 & 1 & 0 &  & 0 & 1 & 0 & 1 & 1 &  & 1 \\ 
			 0 & 0 & 0 & 0 & 1 &  & 0 & 1 & 0 & 0 & 1 &  & 1 \\ 
			 0 & 1 & 0 & 0 & 0 &  & 0 & 0 & 0 & 0 & 1 &  & 1 \\ 
			 0 & 0 & 0 & 0 & 1 &  & 0 & 1 & 0 & 1 & 1 &  & 1 \\ 
			 0 & 0 & 0 & 0 & 1 &  & 0 & 0 & 1 & 1 & 1 &  & 1 \\ 
			 1 & 0 & 0 & 0 & 1 &  & 0 & 0 & 0 & 0 & 1 &  & 1 \\ 
			 0 & 0 & 0 & 1 & 1 &  & 0 & 1 & 0 & 0 & 1 &  & 1 \\ 
			 0 & 0 & 0 & 0 & 0 &  & 0 & 0 & 0 & 1 & 0 &  & 1 \\ 
			 1 & 0 & 0 & 0 & 0 &  & 0 & 1 & 0 & 0 & 1 &  & 1 \\ 
			 0 & 0 & 0 & 1 & 0 &  & 0 & 1 & 0 & 0 & 1 &  & 1 \\ 
			 0 & 0 & 0 & 1 & 0 &  & 0 & 1 & 0 & 1 & 1 &  & 1 \\ 
			 0 & 1 & 0 & 1 & 0 &  & 0 & 1 & 1 & 1 & 1 &  & 1 \\ 
			\bottomrule
		\end{tabular}
	\end{table}

\begin{table}
	\caption{\label{t:popavg} Population averaged tables for all SOC components. `Yes' denotes contact was made with the component, `No' indicates contact was not made. Within each table, the diagonal counts within each table represent the number of concordant pairs while the off-diagonal counts represent the number of discordant pairs.}
	\begin{center}
	\begin{tabular}{rcccrcc}
		 \multicolumn{3}{c}{Development Disabilities} & \hspace{10pt} & \multicolumn{3}{c}{Mental Health} \\
		\cmidrule{1-3} \cmidrule{5-7}
		Specialty & \multicolumn{2}{c}{Primary Care} &  & Specialty & \multicolumn{2}{c}{Primary Care}  \\
		\cmidrule{2-3} \cmidrule{6-7}
		 Care & Yes & No &  & Care & Yes & No  \\
		\cmidrule{1-3} \cmidrule{5-7}
		Yes & $n_{111} = 0$ & $n_{121} = 0$ &  & Yes & $n_{112} = 12$ & $n_{122} = 12$ \\
		No & $n_{211} = 3$ & $n_{221} = 71$ &  & No & $n_{212} = 4$ & $n_{222} = 46$ \\
		\cmidrule{1-3} \cmidrule{5-7}
		\\
	\end{tabular}
	\begin{tabular}{rcccrcc}
		 \multicolumn{3}{c}{Child Welfare} & \hspace{10pt} & \multicolumn{3}{c}{Education} \\
		\cmidrule{1-3} \cmidrule{5-7}
		Specialty & \multicolumn{2}{c}{Primary Care} &  & Specialty & \multicolumn{2}{c}{Primary Care} \\
		\cmidrule{2-3} \cmidrule{6-7}
		 Care & Yes & No &  & Care & Yes & No \\
		\cmidrule{1-3} \cmidrule{5-7}
		Yes & $n_{114} = 3$ & $n_{124} = 7$ &  & Yes & $n_{115} = 11$ & $n_{125} = 31$ \\
		No & $n_{214} = 4$ & $n_{224} = 60$ &  & No & $n_{215} = 2$ & $n_{225} = 30$ \\
		\cmidrule{1-3} \cmidrule{5-7}
		\\
	\end{tabular}
	\begin{tabular}{rcc}
		\multicolumn{3}{c}{Juvenile Justice}\\
		\cmidrule{1-3}
		Specialty & \multicolumn{2}{c}{Primary Care} \\
		\cmidrule{2-3}
		 Care & Yes & No \\
		\cmidrule{1-3}
		Yes & $n_{113} = 1$ & $n_{123} = 2$ \\
		No & $n_{213} = 0$ & $n_{223} = 71$\\
		\cmidrule{1-3}
		\\
	\end{tabular}
	\end{center}
\end{table}

Several authors consider non-Bayesian approaches for analyzing MMP data. Klingenberg and Agresti \cite{Kling2006} use marginal probability models to test for simultaneous marginal heterogeneity. They employ a generalized estimating equation (GEE) with an identity link and working independence to estimate effects and construct a multivariate version of McNemar's test \citep{McNemar1947}. Some examine methods for only pairs of correlated proportions \cite{Consonni2008, Saeki2017}. Others propose multiple testing approaches to account for correlation between sets of matched proportions \citep{Westfall2010,Xu2013}. For example, Westfall, Troendle, and Pennello \cite{Westfall2010} obtain loose control of the family wise error rate using Bonferroni-Holm on bootstrapped $p$-values. Lui and Chang \cite{Lui2013} show that mixed-effects exponential risk models can be used to estimate the risk ratios for each set of matched proportions based on consistent estimators of the risk. In later work, the same authors consider a mixed-effects logistic regression model with separate random effects of pair and set \cite{Lui2016}. The authors show the estimated effect of treatment for each outcome using this approach reduces to the $\log$ of the well-known univariate Cochran-Mantel-Haenszel (CMH) estimator for stratified $2\times 2$ tables \citep{Cochran1950,Mantel1959}, after integrating out the random effects. Jiang and Xu \cite{Jiang2017} explore power and sample size calculations for this and the GEE-based marginal probability model. As is common in the analysis of matched proportions, each of these methods exclude the concordant pairs, relying only on information from the discordant pairs \citep{Agresti2013}.

The literature on Bayesian methods for matched proportions is limited to the univariate case. For example, Altham \cite{Altham1971} examines closed form results for a multinomial model with Dirichlet priors while Broemeling and Gregurich \cite{Broemeling1996} discuss the Gibbs sampling approach to this model. Ghosh et al. \cite{Ghosh2000} present an item response model with logit, probit, and complementary $\log$-$\log$ links. This formulation allows for a hierarchical Bayesian modeling of the data. The authors show that, for such models, the concordant pairs also contribute to estimation and inference. While these approaches are useful for univariate matched proportions, to our knowledge, no previous work considers Bayesian methods for the analysis of MMP data.

More importantly, two of the existing methods for MMP data cannot directly handle sparse responses, as we will now illustrate. The GEE-based approach estimates marginal probabilities using the identity link under the assumption of working independence \citep{Kling2006}. Using a model with one parameter for each marginal probability, the authors show that the estimates equal the sample probabilities of the columns of Table~\ref{t:soc}. But when a column is sparse, as is the case for the developmental disabilities component of the SOC data, the estimated probability is zero. The estimated variance will also be zero, as will the associated covariances. The variance of the difference in marginal probabilities between care-type depends on both this variance and the corresponding covariance of the sparse column, resulting in underestimation of the variance of the difference. The bootstrap-based method pre-differences the paired outcomes, i.e. takes the difference between paired columns of Table~\ref{t:soc}, and estimates the differences in the marginal probabilities using sample averages \citep{Westfall2010}. In general, these differences can take on the values $-1, 0,$ and 1. But when the response is sparse, as in the developmental disabilities component, the differences can only be 0 or 1 meaning the lower bound of a bootstrapped interval is bounded below by 0. Similarly, for the juvenile justice component, only $-1$ and 0 are possible realizations from the data resulting in a bootstrapped interval that is bounded above by 0. Once again, the uncertainty of the difference in marginal probabilities for sparse outcomes is underestimated.

Lui and Chang do both provide adjustments for sparse data in both their exponential risk model (ERM) and their mixed effects regression approach \citep{Lui2013, Lui2016}. The ERM approach, for example, bases its estimation off of the $K$-specific tables in Table~\ref{t:popavg}. Specifically, the estimate is the ratio of the counts of discordant pairs in each table: $n_{21k} \big/ n_{12k}$. When one of these elements is zero, the authors suggest adding $1/2$ to the numerator and the denominator making their sparsity adjusted estimator $\left(n_{21k} + \frac{1}{2}\right) \big/ \left(n_{12k}+ \frac{1}{2}\right)$. Lui and Chang do not evaluate the performance of their estimators in the presence of sparse response and instead evaluate the performance of their testing procedure in non-sparse simulated data settings \citep{Lui2013, Lui2016}. Such a correction is philosophically Bayesian in its thinking as it recognizes, a priori, that the true population parameter is likely not zero (or infinite) and adjusts the estimate accordingly. In fact, this estimator is an empirical Bayes estimate of the marginal risk for a single matched set under Jeffreys' prior when using the Dirichlet model proposed by \cite{Altham1971}.

Adjustments of this kind are not generalizable either: it is not applicable to the GEE or Bootstrap approaches as it adjusts the counts from Table~\ref{t:popavg} whereas both the GEE and Bootstrap rely on the individual-level data as in Table~\ref{t:soc}. Other common adjustments like data augmentation, e.g. adding one success and one failure in the case of binary data, do not easily generalize to the multivariate case either. Augmenting sparse cells in Table~\ref{t:popavg} separately ignores the multivariate nature of the data, breaking potential underlying correlations. Adding covariate patterns to Table~\ref{t:soc} must take into account the impact on the multivariate structure and can lead to severe sample size inflation. At the extreme, augmentation would involve adding one of every possible type of covariate pattern to Table~\ref{t:soc}. For even a small number of sets, such an addition would greatly inflate the sample size far beyond what augmentation strategies aim for, potentially biasing the results. Thus, there is a need for an MMP method that can accommodate sparse responses more adequately in addition to the need for a Bayesian approach.

Using the Albert and Chib \cite{AlbertChib1993,AlbertChib1995} probit model as a basis, we consider several approaches for the Bayesian modeling of multivariate matched proportions. The latent variable representation of the probit model solves the issue of variance estimation in the existing methods: by convention, the latent-space variance is fixed at one. This alone may not be sufficient in the presence of sparse response where estimates can easily diverge, so we also consider shrinkage priors on model components. Finally, noting that the first two approaches assume independence in the latent space, we examine a multivariate analysis. We refer to the three Bayesian models we examine as the na\"ive analysis which uses a simple Bayesian probit model, the penalized analysis which uses half-Cauchy priors on the variance components of a global ridge-like prior \cite{Gelman2006}, and the multivariate analysis. Our multivariate analysis makes use of a Bayesian functional principal components analysis or FPCA \cite{VanDerLinde2008} to model a general latent covariance matrix. The Bayesian FPCA can accommodate a wide range of covariance structures, not just those arising from functional data, making it an appealing approach for dealing MMP data.

In this manuscript, we demonstrate that all Bayesian models have better coverage and power in the presence of sparse responses than the existing methods for MMP data while maintaining similar bias and interval widths. Comparing between the Bayesian models, we find that the multivariate analysis tends to perform best in terms of coverage, although the na\"ive and penalized analyses do perform similarly well. Finally, we reanalyze the SOC data using the multivariate analysis and include the two components with sparse responses which were excluded from the original analysis. Our reanalysis confirms some of the original results but also identifies additional findings the univariate analysis missed.

\section{Methods}
\label{s:methods}

We begin by defining an overarching notation for the models we consider. At the pair-level, let $x_{ijk}$ denote the response for pair $i$, at observation $j$, for the $k$th set where $i = 1, \ldots, n$, $j = 1, 2$, and $k = 1, \ldots, K$.  In context of the SOC data, $i$ denotes the subjects and $j$ denotes care-type, primary care or specialty care. The index $k$ denotes the SOC component with which contact is made, $x_{ijk} = 1$, or not made,  $x_{ijk} = 0$. Combining across $k$, we let $\bx_{ij} = \left[ \begin{array}{ccc} x_{ij1} & \cdots & x_{ijK} \end{array} \right]'$ be the vector of responses at observation $j$ for pair $i$. Subject $i$'s stacked vector of responses is then $\bx_i = \left[ \begin{array}{cc} \bx_{i1} & \bx_{i2} \end{array} \right]'$. Realizations of this vector make up the rows of Table~\ref{t:soc}.

The target of inference is a vector consisting of the differences in the $k$-specific marginal probabilities of contact, $P(x_{ijk} = 1) = \theta_{jk}$, which both the GEE-based \cite{Kling2006} and bootstrap-based \cite{Westfall2010} methods estimate. First let $\btheta = \left[ \begin{array}{cccccc} \theta_{11} & \cdots & \theta_{1K} & \theta_{21} & \cdots & \theta_{2K} \end{array} \right]$ be the vector of marginal probabilities. Then let $\rho_k$ be the $k$th difference in the marginal probabilities, thus $\rho_k = \theta_{1k} - \theta_{2k}$. The vector of $\rho_k$'s is $\brho$,  $\brho = \left[ \begin{array}{ccc} \rho_{1} & \cdots & \rho_{K} \end{array} \right]$. Alternatively, define $\bL$ to be a block matrix of the form $\bL = \left( \bI_{K} \vert  -\bI_{K} \right)$ where $\bI_{K}$ is a $K \times K$ identity matrix. The vector representing the differences in marginal probabilities is then $\brho = \bL\btheta$.

\subsection{Na\"ive Analysis}
\label{s:naive}

We first consider a na\"ive analysis using probit regression as first described by Albert and Chib \cite{AlbertChib1993,AlbertChib1995}. The regression framework makes use of the latent variable representation of the probit model. In the Bayesian context, this approach can be considered a data augmentation step where the latent variables are the augmented data \cite{Gelman2013}. We define the following mapping
\begin{align}
	x_{ijk} = \left\{\begin{array}{cc}
				0 & \text{ if  }z_{ijk} < 0\\
				1 & \text{ if  }z_{ijk} \geq 0
			\end{array}\right.,\label{eq:lv}
\end{align}
where the latent space model for MMP data is $z_{ijk} = \beta_{jk} + \epsilon_{ijk}$. In general, $\epsilon_{ijk} \sim N(0, \sigma_z^2)$ and, by convention, $\sigma_z^2$ is fixed at 1. Because the model is saturated, the relationship between the latent space coefficients, $\beta_{jk}$, and the marginal probabilities, $\theta_{jk}$, is $\beta_{jk} = \Phi^{-1}(\theta_{jk})$ where $\Phi(\cdot)$ is the standard normal CDF. This relationship, and indeed the probit model itself, is induced by assuming the error terms are normal. Letting $\bbeta = \left[ \begin{array}{cccccc} \beta_{11} & \cdots & \beta_{1K} & \beta_{21} & \cdots & \beta_{2K} \end{array} \right]$, we place the a flat, non-informative prior on $\bbeta$, specifically $\pi(\bbeta) \propto 1$. The resulting posterior conditionals are:
\begin{align*}
	z_{ijk} &\vert x_{ijk} = 0, \beta_{jk} \sim N(\beta_{jk}, 1)_{\{z_{ijk} < 0\}}\\
	z_{ijk} &\vert x_{ijk} = 1, \beta_{jk} \sim N(\beta_{jk}, 1)_{\{z_{ijk} \geq 0\}}\\
	\bbeta &\vert \bZ, \bW \sim N\left[ (\bW'\bW)^{-1} \bW'\bZ, (\bW'\bW)^{-1} \right]
\end{align*}
where $\bW$ is a design matrix built by stacking $2K\times2K$ identity matrices on top of each other, one for each subject, and $\bZ$ is the vector of all $z_{ijk}$'s. The subscript notation $\{ \}$ on the normal densities indicates truncation to the region defined in the brackets. Estimates of the difference in marginal probabilities, $\brho$, are obtained from the transformed posterior samples of $\bbeta$: $\brho = \bL\Phi(\bbeta)$ where the function $\Phi(\cdot)$ is applied to the vector $\bbeta$ in component-wise fashion.

In the above conditionals, $\sigma_z^2$ is fixed at one. Since the latent variable representation is scale-invariant, $\sigma_z^2$ is not necessarily identifiable. Fixing $\sigma_z^2$ at one solves the identifiability issue in the latent space. But this convention provides an additional benefit in the presence of sparse response by stabilizing the model variance. Without this constraint, the posterior chains for $\beta_{jk}$ and, subsequently, $\rho_k$, would not converge in the presence of sparse responses. Even with this constraint, the posterior chains can be unstable because in the latent space, the support of each $\beta_{jk}$ is $\mathbb{R}$. To prevent the model components from diverging, we penalize the $\beta_{jk}$ and shrink them toward zero.

%
%

\subsection{Penalized Analysis}

There are many choices for shrinkage options in the Bayesian framework, but we select normal priors with a half-Cauchy hyper-prior on the scale. Gelman \cite{Gelman2006} proposed the use of half-$t$ priors for variance components, of which the half-Cauchy is a special case. Polson and Scott \cite{Polson2012} show that the half-Cauchy performs well as global-shrinkage prior and recommend their routine use in normal hierarchical models. The model specification for the penalized analysis begins in the same fashion with the mapping in Equation~\eqref{eq:lv} and with the same latent model:  $z_{ijk} = \beta_{jk} + \epsilon_{ijk}$ for $\epsilon_{ijk} \sim N(0, 1)$. Instead of the flat prior, we place the following global hierarchical prior on $\beta_{jk}$:
\begin{align}
	\beta_{jk} &\stackrel{iid}{\sim} N(0, \lambda)\nonumber\\
	\lambda &\sim IG\left( \frac{1}{2}, \frac{1}{\mu} \right)\label{eq:betapen}\\
	\mu &\sim IG\left( \frac{1}{2}, \frac{1}{A^2} \right)\nonumber
\end{align}
where $IG$ denotes the inverse gamma distribution and $A$ is a fixed hyper-parameter. These hierarchical inverse gamma priors are the mixture representation of the half-Cauchy prior \cite{Wand2011}. That is, if $\lambda \sim IG\left( 1/2, 1/\mu \right)$ and $\mu \sim IG\left( 1/2, 1/A^2 \right)$, then $\sqrt{\lambda} \sim HC(A)$ where $HC$ denotes the half-Cauchy.

The mixture-model representation of the half-Cauchy induces a model with fully identifiable conditional posterior distributions:
\begin{align*}
	z_{ijk} &\vert x_{ijk} = 0, \beta_{jk} \sim N(\beta_{jk}, 1)_{\{z_{ijk} < 0\}}\\
	z_{ijk} &\vert x_{ijk} = 1, \beta_{jk} \sim N(\beta_{jk}, 1)_{\{z_{ijk} \geq 0\}}\\
	\bbeta &\vert Z, \bW \sim N\left[ (\bW'\bW + \lambda^{-1} \bI_{2K})^{-1} \bW'\bZ, (\bW'\bW +  \lambda^{-1} \bI_{2K})^{-1} \right]\\
	\lambda &\vert \bbeta, \mu  \sim IG\left(K + \frac{1}{2}, \frac{1}{\mu} + \frac{1}{2} \bbeta'\bbeta \right)\\
	\mu &\vert \lambda  \sim IG\left( 1, \frac{1}{A^2} + \frac{1}{\lambda} \right)
\end{align*}
where $\bI_{2K}$ is the $2K \times 2K$ identity matrix, $\bW$ is as defined in Section~\ref{s:naive}. This model provides global shrinkage of the latent-space coefficients, $\beta_{jk}$, across all $j$ and $k$ and stabilizes the chain when the response is sparse. Estimates of $\brho$ are once again obtained from the transformed samples of $\bbeta$ using $\brho = \bL\Phi(\bbeta)$. 

The choice of the hyper-parameter $A$ can be regarding as placing a weak upper bound on the variance of $\bbeta$ \cite{Gelman2006}. The parameter can be interpreted in context as the standard deviation of the corresponding model component \cite{Gelman2006}. Since we are modeling in a latent space, our choice of the value of $A$ is less clear as the scale is not directly dependent on a covariate. Thus, we select a relatively large value, $A = 10$, to give a diffuse prior.


%
%

\subsection{Multivariate Analysis}

Both of the preceding methods treat the data as independent univariate samples, however each subject's $\bx_i$ may have a multivariate distribution. Chib and Greenberg \cite{ChibGreenberg1998} discuss a multivariate version of the Albert and Chib \cite{AlbertChib1993,AlbertChib1995} model which, in the MMP context, begins with the same mapping as in Equation~\eqref{eq:lv}. The latent model is then $\bz_i = \bbeta + \beps_i$ where $\bz_i = \left[ \begin{array}{cccccc} z_{i11} & \cdots & z_{i1K} & z_{i21} & \cdots & z_{i2K} \end{array} \right]'$ and $\beps_i = \left[ \begin{array}{cccccc} \epsilon_{i11} & \cdots & \epsilon_{i1K} & \epsilon_{i21} & \cdots & \epsilon_{i2K} \end{array} \right]'$. The latent error vector, $\beps_i$, is then assumed to be multivariate normal, $\beps_i \sim N(\zero, \bSigma)$. The matrix $\bSigma$ is a correlation matrix.

Drawing posterior samples for $\bSigma$ can be difficult. Chib and Greenberg suggest a Metropolis-Hastings step to obtain draws from the posterior of $\bSigma$ and others authors have further discussed the same issue \cite{Liu2001,Zhang2006,Webb2008}. In all of these approaches, the MCMC sampler also requires obtaining draws from a truncated multivariate normal posterior, the posterior conditional distribution of $\bz_i$, for each subject. Using existing approaches for random drawing from truncated multivariate normal distributions is time intensive. In our preliminary model exploration, a model of this kind took over 11 hours to generate 20,000 posterior samples when $K = 2$ (run on a computer with a 3.7 GHz 6-Core Intel Core i5 processor and 32 GB of memory).





To accommodate a multivariate structure without inducing such a computational burden, we turn to recent developments in the functional data literature. In particular, Goldsmith and Kitago \cite{Goldsmith2016} and Meyer et al. \cite{MeyerMorris2022} both show that the Bayesian FPCA \cite{VanDerLinde2008}, can be used to model a variety of covariance structures for both Gaussian \cite{Goldsmith2016} and ordinal \cite{MeyerMorris2022} functional data including independent, exponential, and compound symmetric (also known as exchangeable) covariance structures. For MMP data, we may expect exchangeability within observation level $j$, what might be called a block compound symmetric structure, or independence within $j$ but dependence between $k$. A more general form or unstructured form may result as well. An additional challenge is that it is hard to know what the structure of the latent correlation matrix might be. Thus, for the multivariate analysis, we propose using an FPCA in the latent space to model possible multivariate associations in MMP data.

Our proposed model performs the FPCA on the latent errors setting $\beps_i = \bOmega\bpsi\bc_i' + \bVareps_i$ for the $L\times 1$ vector of subject scores $\bc_i$, the $2K\times L$ matrix of basis coefficients $\bPsi$, and the $2K\times2K$ matrix of basis-functions $\bOmega$. The model for each subject's latent vector is then
\begin{align*}
	\bz_i = \bbeta + \bOmega\bPsi\bc_i' + \bVareps_i,
\end{align*}
where $\bVareps_i$ is a $2K\times1$ vector of normal, independent errors with variance equal to $\sigma_{\varepsilon}^2\bI_{2K}$. Consistent with previous work using Bayesian FPCA, we use $L = 2$ scores for the expansion \cite{Goldsmith2016, MeyerMorris2022}. Letting $\bOmega$ be a matrix of cubic B-spline basis functions, the prior on $\bPsi$ is then $\bPsi \sim N[0, (\bLambda_{\psi}\otimes\pmat)^{-1}]$ where $\bLambda_{\psi}$ is an $L\times L$ diagonal matrix of tuning parameters $\lambda_{\ell}^{-1}$ and $\pmat$ is a $2K\times2K$ penalty matrix of the form $\pmat = \xi \pmat_0 + (1 - \xi)\pmat_2$. The matrices $\pmat_0$ and $\pmat_2$ are the zeroth and second derivative penalty matrices, respectively. The value $\xi$ is a control parameter that balances smoothness and shrinkage with values near zero favoring shrinkage---we take $\xi = 0.01$. Finally, the subject scores, $\bc_i$, have mean zero normal priors with variance equal to $\bI_{L}$, an $L \times L$ identity matrix.

We use the same prior on $\bbeta$ as in the penalized analysis, found in Equation~\eqref{eq:betapen}. Combining with the FPCA specification, we can obtain the full conditional posterior distributions for each model component. Let $\Psi_{ijk}$ denote the $j,k$th sub-components of the vector $\bOmega\bPsi\bc_i'$. The full conditionals are
\begin{align*}
	z_{ijk} &\vert x_{ijk} = 0, \beta_{jk}, \Psi_{ijk} \sim N(\beta_{jk} + \Psi_{ijk}, 1)_{\{z_{ijk} < 0\}}\\
	z_{ijk} &\vert x_{ijk} = 1, \beta_{jk}, \Psi_{ijk} \sim N(\beta_{jk} + \Psi_{ijk}, 1)_{\{z_{ijk} \geq 0\}}\\
	\bbeta &\vert Z, \bW \sim N\left[\bSigma_{\bbeta} \bW'(\bZ - \bpsi), \bSigma_{\bbeta} \right]\\
	\lambda &\vert \bbeta, \mu  \sim IG\left(K + \frac{1}{2}, \frac{1}{\mu} + \frac{1}{2} \bbeta'\bbeta \right)\\
	\mu &\vert \lambda  \sim IG\left( 1, \frac{1}{A^2} + \frac{1}{\lambda} \right)\\
	\bPsi &\vert \sim N\left[ \frac{1}{\sigma_{\varepsilon}^2}\bSigma_{\bPsi} (\bC \otimes \bOmega)'(\bZ - \bW\bbeta) , \bSigma_{\bPsi}\right]\\
	\bc_i &\vert \sim N\left[ \frac{1}{\sigma_{\varepsilon}^2}\bSigma_{\bc}(\bOmega\bPsi)'(\bz_i - \bbeta) , \bSigma_{\bc}\right]\\
	\sigma_{\varepsilon}^2 &\vert \sim IG\left[ 1 + nK, 1 + \frac{1}{2}(\bZ - \bW\bbeta - \bpsi)'(\bZ - \bW\bbeta - \bpsi) \right]\\
	\lambda_{\ell} &\vert \sim IG\left(2K, K + \frac{1}{2} \bPsi_{\ell}' \pmat \bPsi_{\ell}  \right)
\end{align*}
for $\bSigma_{\bbeta} = \left(\frac{1}{\sigma_{\varepsilon}^2}\bW'\bW +  \lambda^{-1} \bI_{2K}\right)^{-1}$, $\bSigma_{\bPsi} =\left( \frac{1}{\sigma_{\varepsilon}^2} \bC'\bC \otimes \bOmega'\bOmega + \bLambda_{\psi}\otimes\pmat \right)^{-1}$, $\bSigma_{\bc} = \left[\frac{1}{\sigma_{\varepsilon}^2}(\bOmega\bPsi)'(\bOmega\bPsi) + \bI_{L} \right]^{-1}$, $\bC$ equal to the matrix of $\bc_i$ vectors stacked on top of each other, $\bPsi_{\ell}$ equal to the $\ell$th column of $\bPsi$, and $\bpsi$ the vector containing all $\Psi_{ijk}$s. Note that we draw samples from $\sigma_{\varepsilon}^2$ to improve the estimation of other model components, but fix $\sigma_{\varepsilon}^2$ at one when sampling the latent variable.

This expansion allows us to treat the latent variables as we would in the univariate analysis and does not require sampling from a multivariate truncated normal nor does it necessitate a Metropolis-Hastings step to draw latent correlations. As with the na\"ive and penalized analyses, estimates of $\brho$ are easily obtained from the posterior of $\bbeta$ with $\brho = \bL\Phi(\bbeta)$. This model is much more computationally efficient than other multivariate probit models taking between 60 and 70 seconds, depending on $K$, to generate 20,000 posterior samples (on the same computer, one with a 3.7 GHz 6-Core Intel Core i5 processor and 32 GB of memory).

\section{Simulation Study}
\label{s:sim}

We base our simulation study off of the sample size of and effect sizes observed in the SOC data. For each dataset, we generate data for $n = 75$ subjects where $K = 2, 3, 4,$ and 5 sets of binary measurements are taken on each subject. We aim to induce sparse responses within sets and achieve this by first generating probabilities for $\btheta$ from a multivariate normal distribution. In the event a draw from this distribution is negative, it is set to zero. Values of $\bx_i$ are then drawn from a Bernoulli distribution. The ``true'' values of $\btheta$ vary with $K$:
\begin{align*}
	K = 2 &: \btheta = \left[\begin{array}{cccc} 0.05 & \theta_{12} & 0.25 & 0.005 \end{array}\right],\\
	K = 3 &: \btheta = \left[\begin{array}{cccccc} 0.05 & \theta_{12} & 0.1 & 0.25 & 0.005 & 0.15 \end{array}\right],\\
	K = 4 &: \btheta = \left[\begin{array}{cccccccc} 0.05 & \theta_{12} & 0.005 & 0.1 & 0.25 & 0.005 & 0.05 & 0.15 \end{array}\right], \text{ and}\\
	K = 5 &: \btheta = \left[\begin{array}{cccccccccc} 0.05 & \theta_{12} & 0.005 & 0.1 & 0.25 & 0.25 & 0.005 & 0.05 & 0.15 & 0.35\end{array}\right].
\end{align*}
The $k = 2$ component is set to exhibit sparse response since $\theta_{22} = 0.005$. We add an additional sparse responses for $K = 4$ and 5, although sparse responses may occur for any component in any given simulated draw. To evaluate the effect of sparse response as the degree of difference between observations changes, we vary $\theta_{12}$ from 0.05 to 0.2 by 0.1. All of these values for $\btheta$ encompass the range of values we see when estimating $\brho$ for the SOC data.

To evaluate the MMP methods, we determine the coverage, power, bias, and interval width for the $k = 2$ component, $\rho_2 = \theta_{21} - \theta_{22}$, which is simulated to exhibit sparse response. We compare all three Bayesian models from Section~\ref{s:methods} to the GEE-based approach \citep{Kling2006}, the Bootstrap-based method \citep{Westfall2010}, and a modified ERM approach \citep{Lui2013}. The ERM estimates risk ratios  \cite{Lui2013}, however, for comparison with the GEE, Bootstrap, and the Bayesian approaches, we estimate the risk difference using their model. All Bayesian estimates are based on 20,000 total samples, discarding the first 10,000 while all Bootstrap estimates are based on 10,000 resamples. Additional details on the GEE, ERM, and Bootstrap approaches are in the Section 1 of the Supplementary Materials.

For each combination of $K$ and $\theta_{12}$, we generate 1000 simulated datasets, retaining only those datasets that exhibit sparsity for the second set---as determined by the sparsity exhibited in either Table~\ref{t:soc} or~\ref{t:popavg}. The parameters of the simulation result in between 294 and 422 datasets exhibiting sparse responses (out of 1000), depending on the value of both $\theta_{12}$ and $K$. We then randomly sample 200 of these datasets for evaluation. All Bayesian models perform similarly in terms of bias. The multivariate model has higher coverage for most values of $\theta_{12}$ and the power advantage of the simpler methods decreases as $K$ increases (see Section 2 of the Supplementary Materials). For these reasons and to ensure we capture the multivariate nature of the data, we suggest using the multivariate analysis.

\begin{figure}[h]%
\centering
\includegraphics[width=0.496\textwidth]{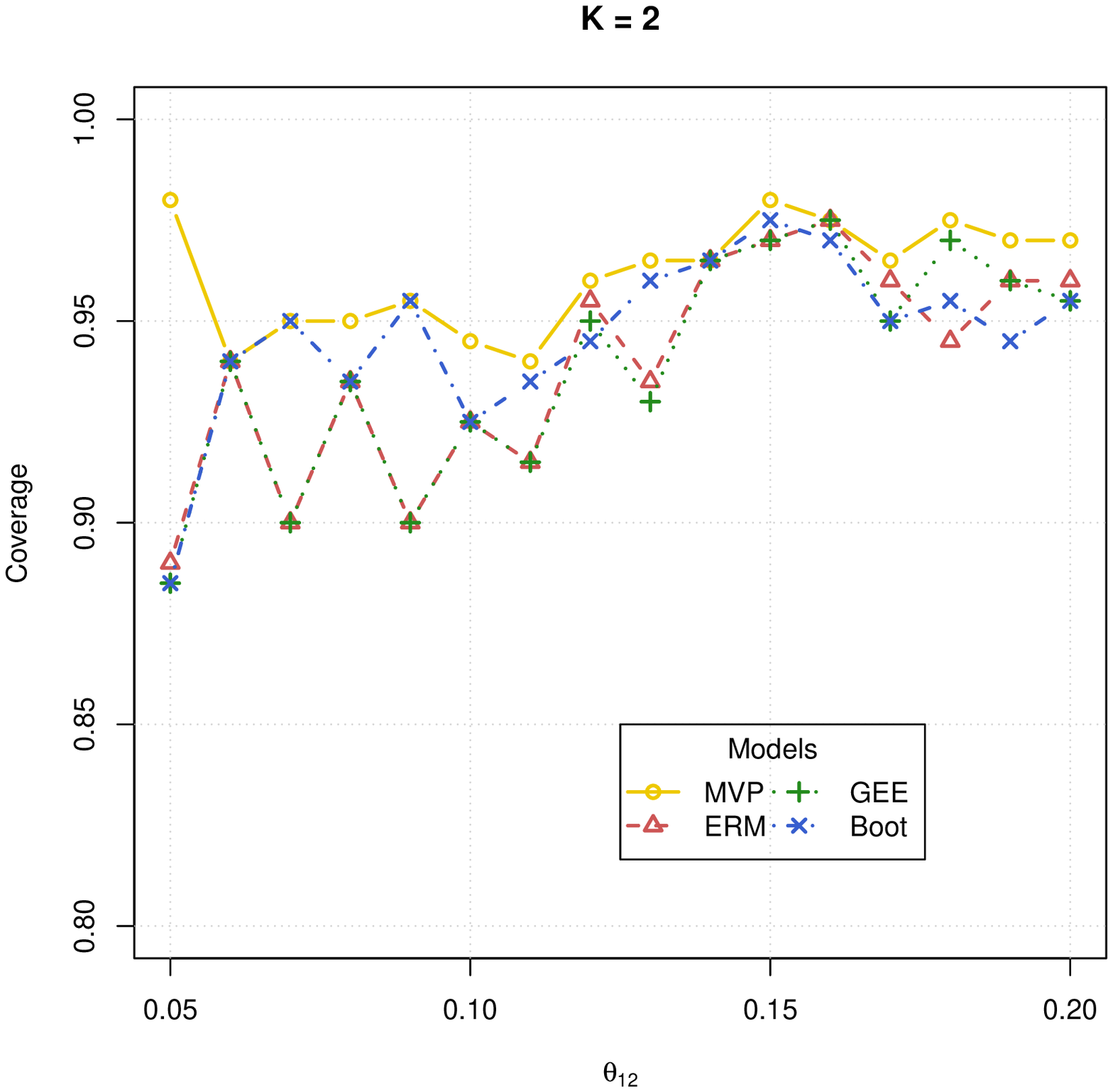}
\includegraphics[width=0.496\textwidth]{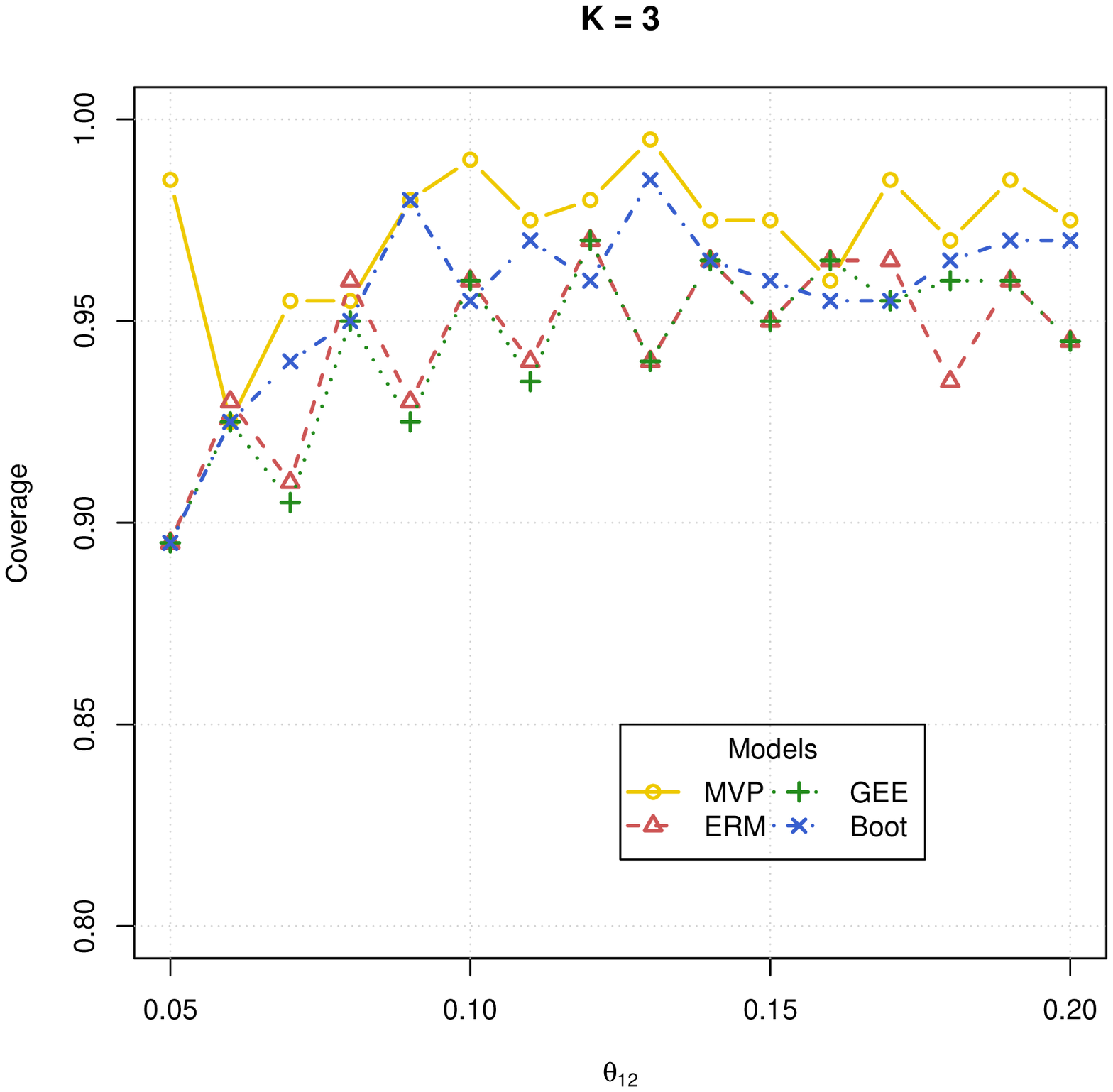}
\includegraphics[width=0.496\textwidth]{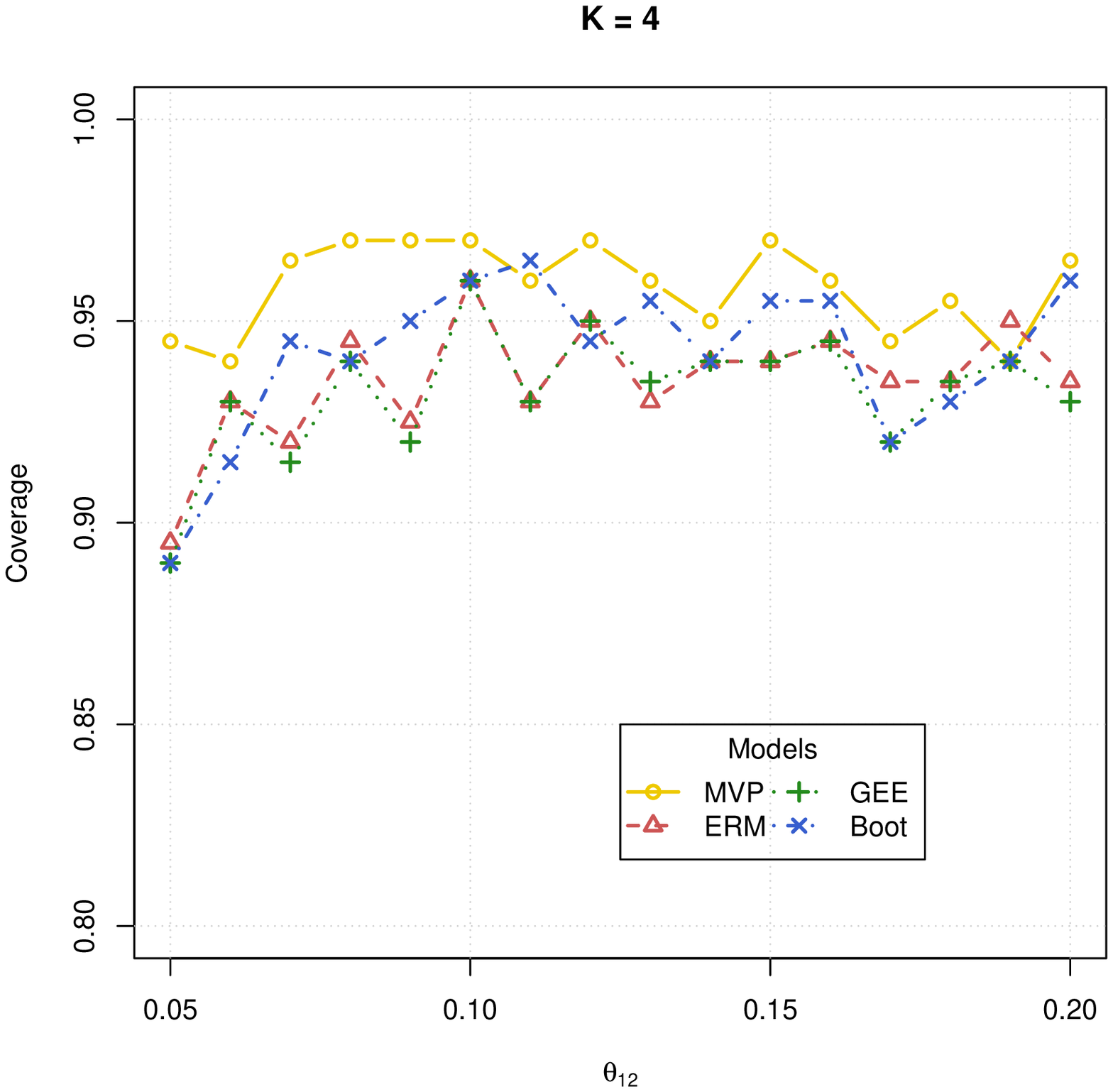}
\includegraphics[width=0.496\textwidth]{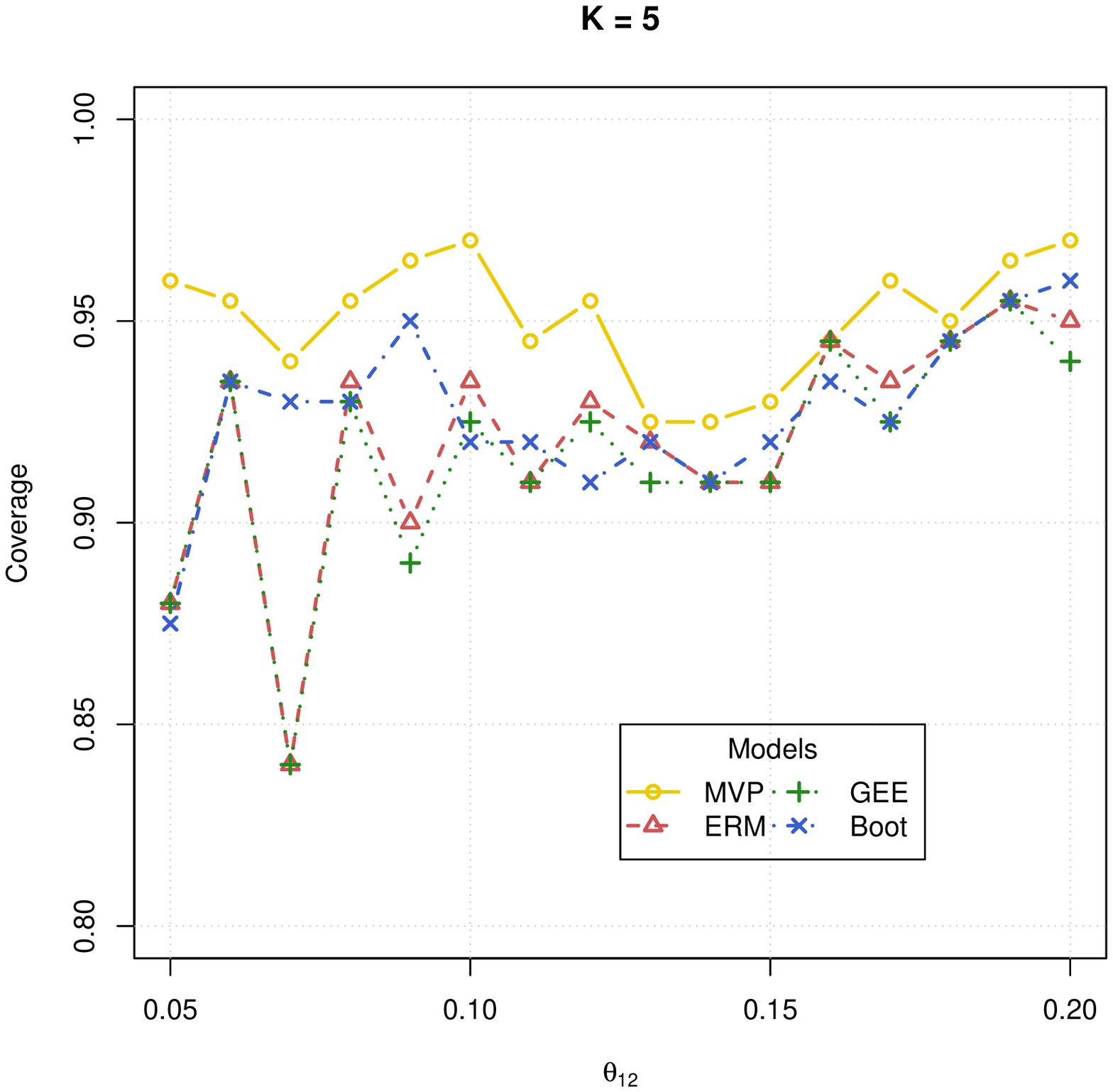}
\caption{Coverage of $\rho_2$ for the multivariate probit analysis (MVP), exponential risk model (ERM), GEE-based model (GEE), and the Bootstrap-based model (Boot) by $\theta_{12}$ and $K$. All intervals are 95\% intervals.}\label{f:cover}
\end{figure}

Figure~\ref{f:cover} contains the coverage comparison between the four models under comparison for all values of $\theta_{12}$ and $K$. Coverage curves for the multivariate probit analysis (MVP) are in solid gold with circles, the exponential risk model (ERM) results are in dashed red with triangles, the GEE-based model (GEE) results are in dotted green with plus signs, and the Bootstrap-based model (Boot) results are in dotted-dashed blue with x's. Consistent with other evaluations of intervals for binary data,\cite{AgrestiCoull1998,AgrestiCaffo2000} we find that the coverage for all methods varies by effect size. The multivariate analysis has the highest coverage for almost every combination of $\theta_{12}$ and $K$, is typically at or above nominal (95\%), and never drops below 92.5\%. The comparator methods, however, have coverage that is frequently below nominal and, in some case, at or below 90\%. Coverage is more inconsistent for the ERM and GEE, particularly when $\theta_{12}$ is smaller. As $\theta_{12}$ increases, all methods have coverage near nominal.

\begin{figure}[h]%
\centering
\includegraphics[width=0.496\textwidth]{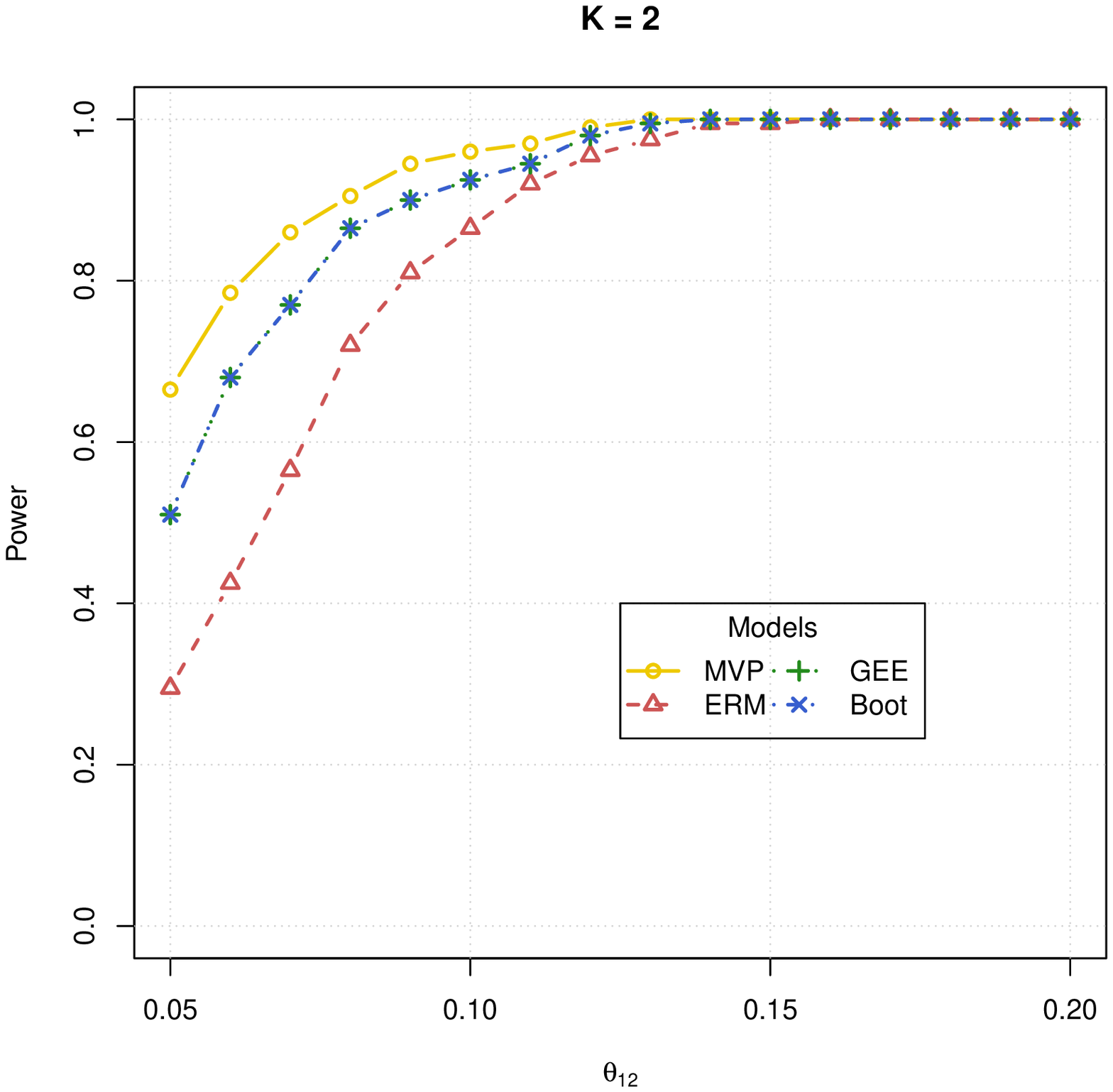}
\includegraphics[width=0.496\textwidth]{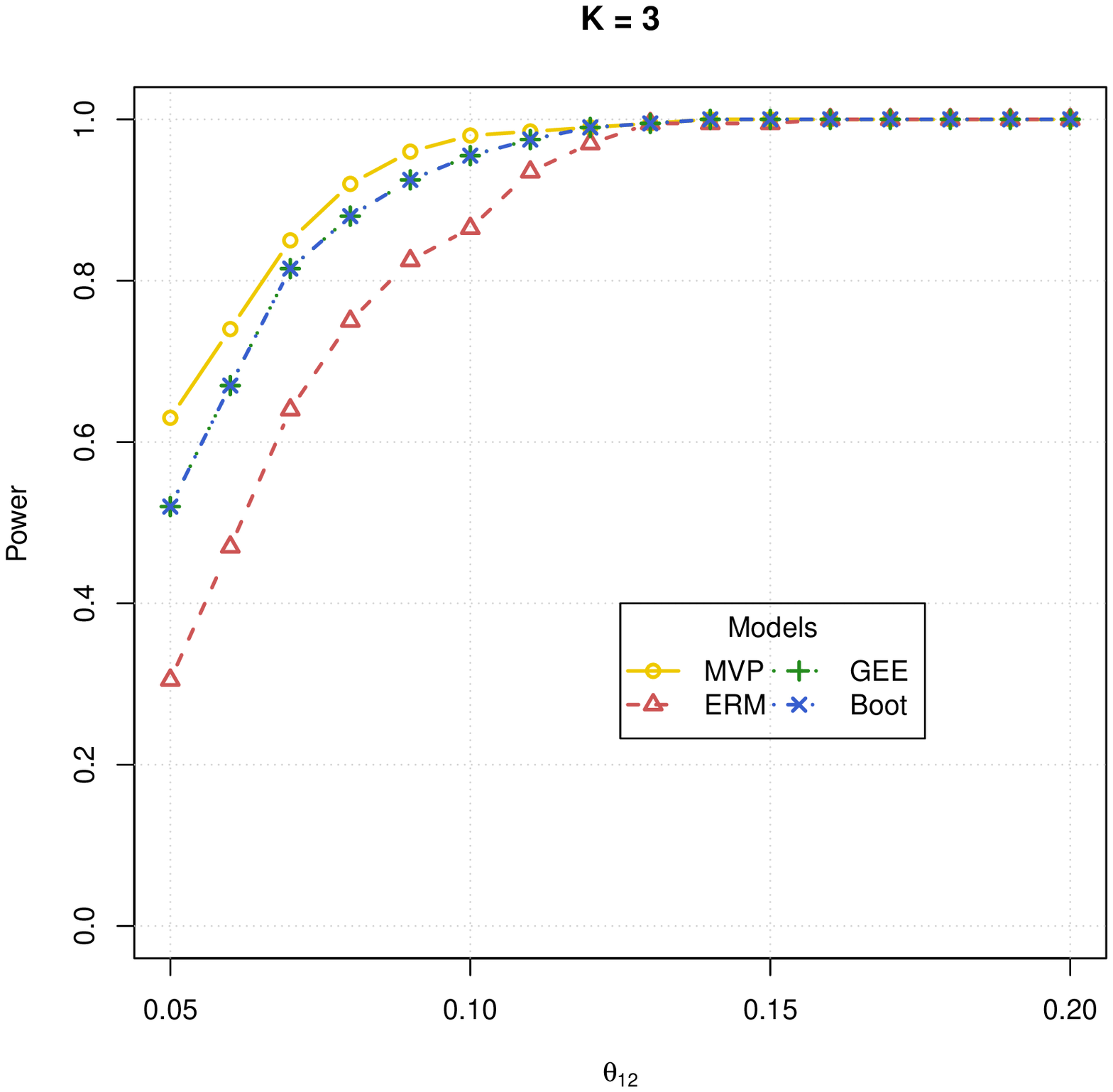}
\includegraphics[width=0.496\textwidth]{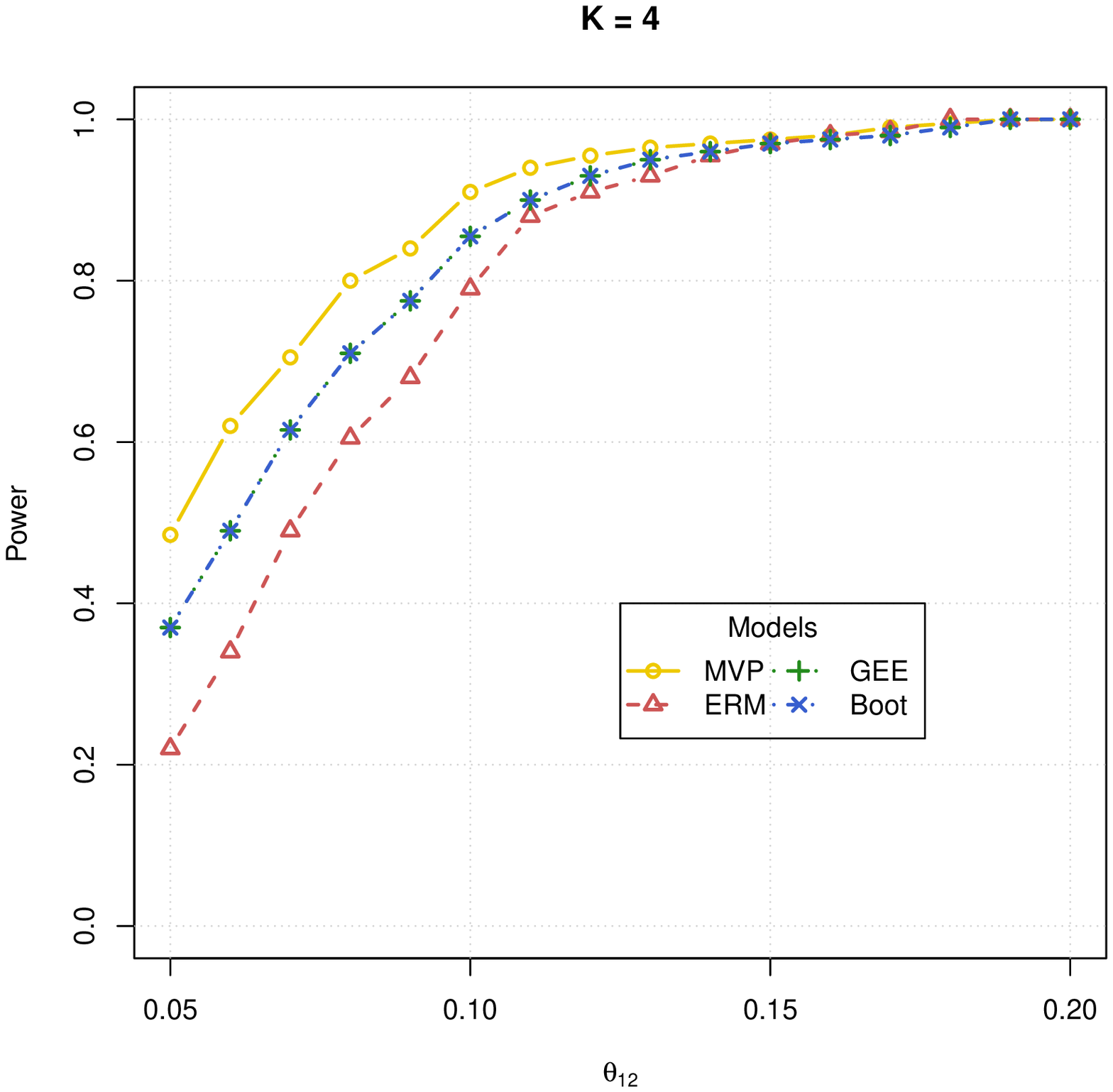}
\includegraphics[width=0.496\textwidth]{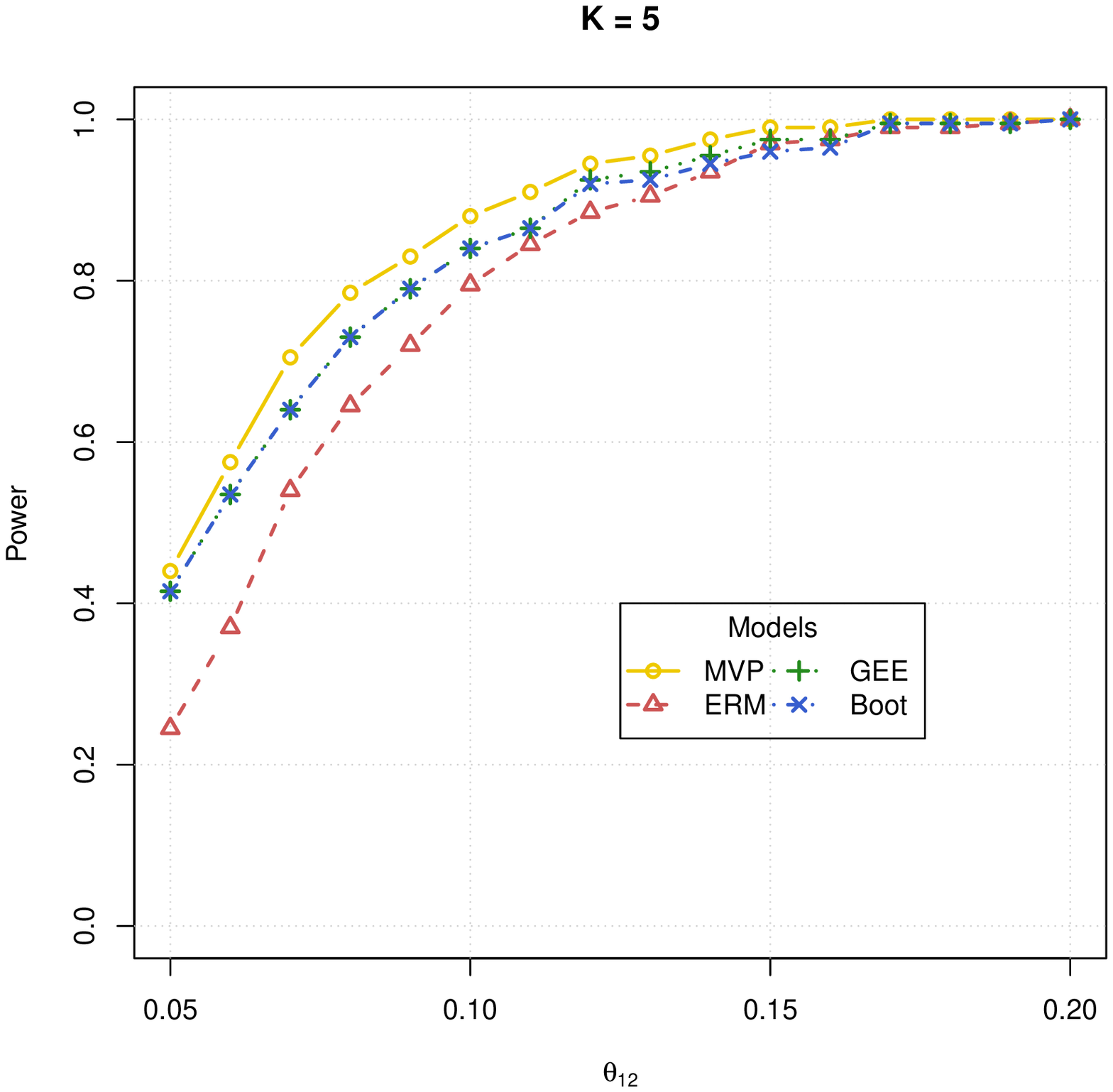}
\caption{Power to detect $\rho_2$ for the multivariate probit analysis (MVP), exponential risk model (ERM), GEE-based model (GEE), and the Bootstrap-based model (Boot) by $\theta_{12}$ and $K$. All tests were conducted at the 5\% level.}\label{f:power}
\end{figure}

A gain in coverage can come at a loss in power, so we also investigate the power of each method to test the null $H_0: \rho_2 = 0$. The results are in Figure~\ref{f:power} where significance is determined by exclusion from the 95\% interval for the multivariate analysis and Bootstrap-based model and by the resulting $p$-values being less than 0.05 for the GEE and ERM. Power curves for the models follow the same color, line, and point schemes as in the depiction of the coverage curves. The multivariate analysis has the highest power at each $\theta_{12}$, regardless of the value of $K$. The GEE and Bootstrap are second with nearly identical power while the ERM has the lowest power. All models have power that goes to one as $\theta_{12}$ increases.

\begin{figure}[h]%
\centering
\includegraphics[width=0.496\textwidth]{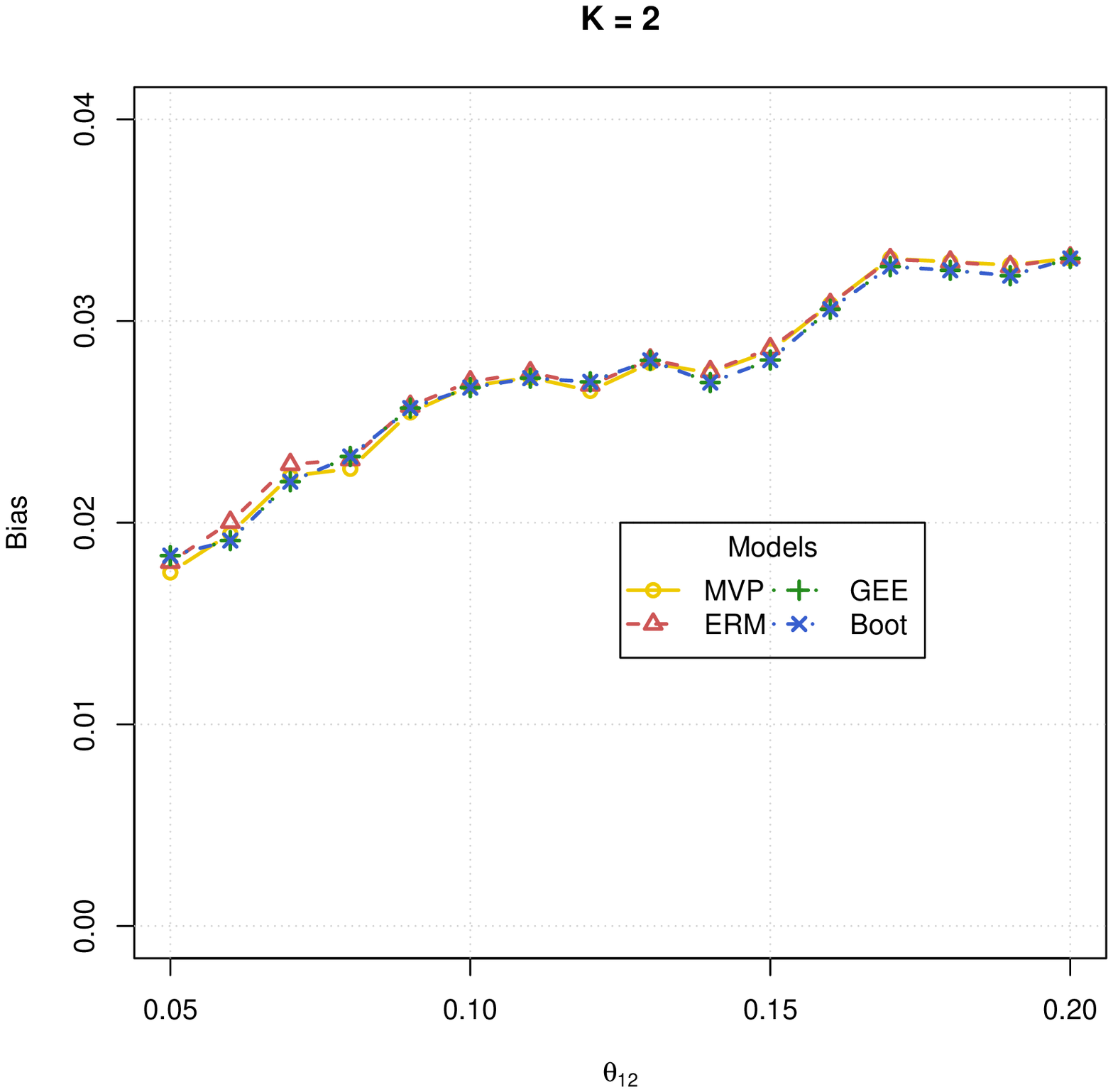}
\includegraphics[width=0.496\textwidth]{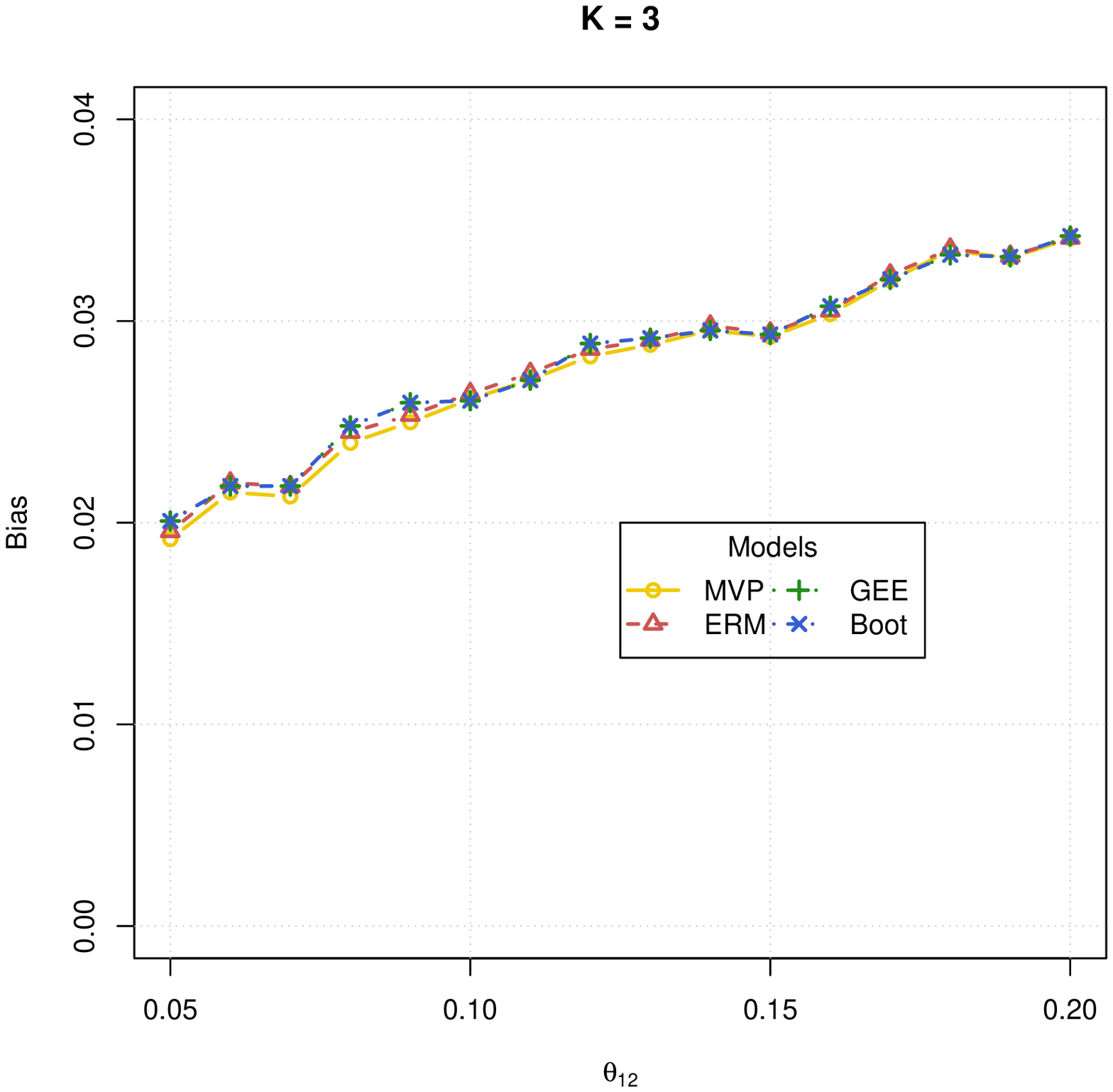}
\includegraphics[width=0.496\textwidth]{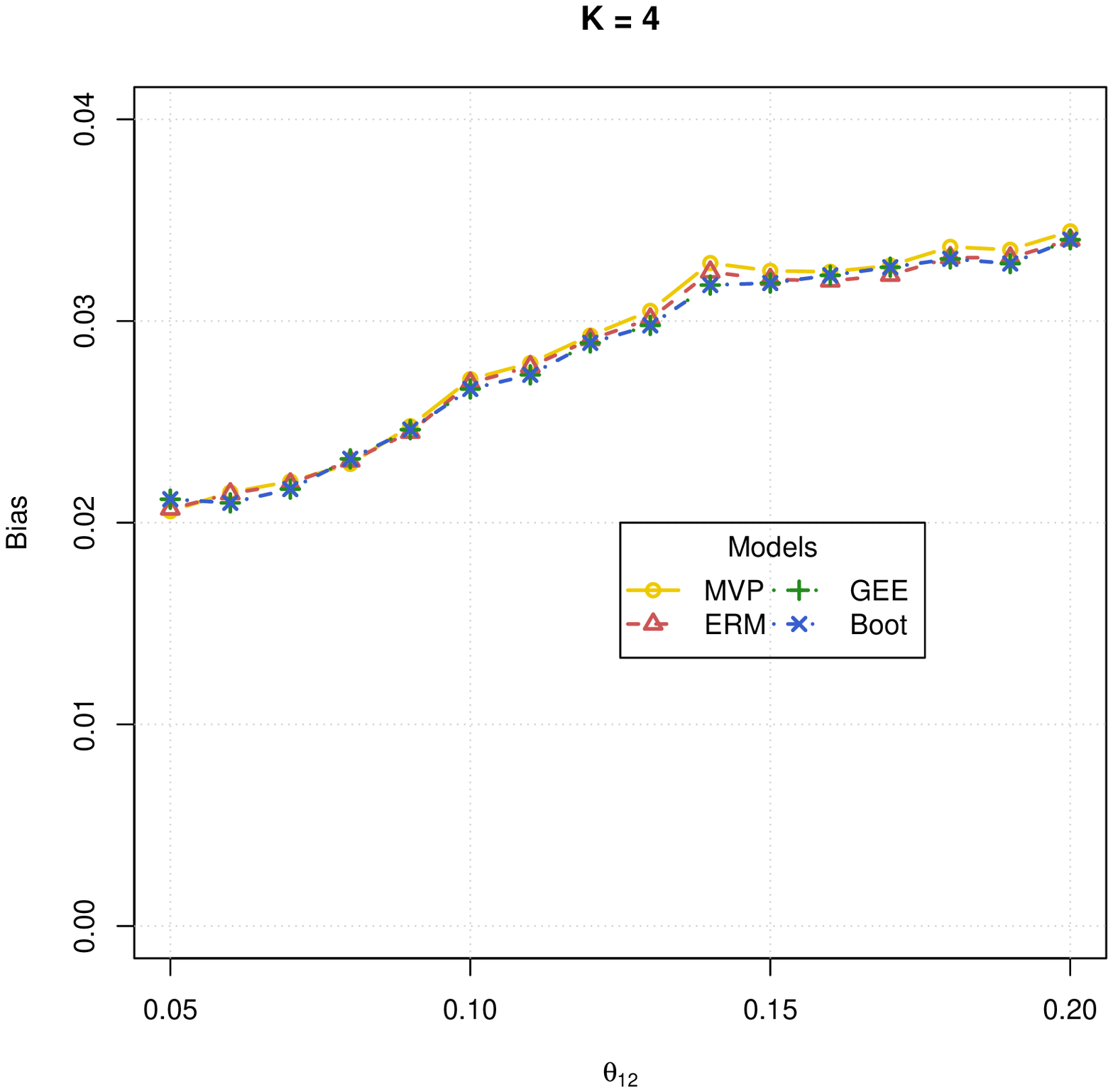}
\includegraphics[width=0.496\textwidth]{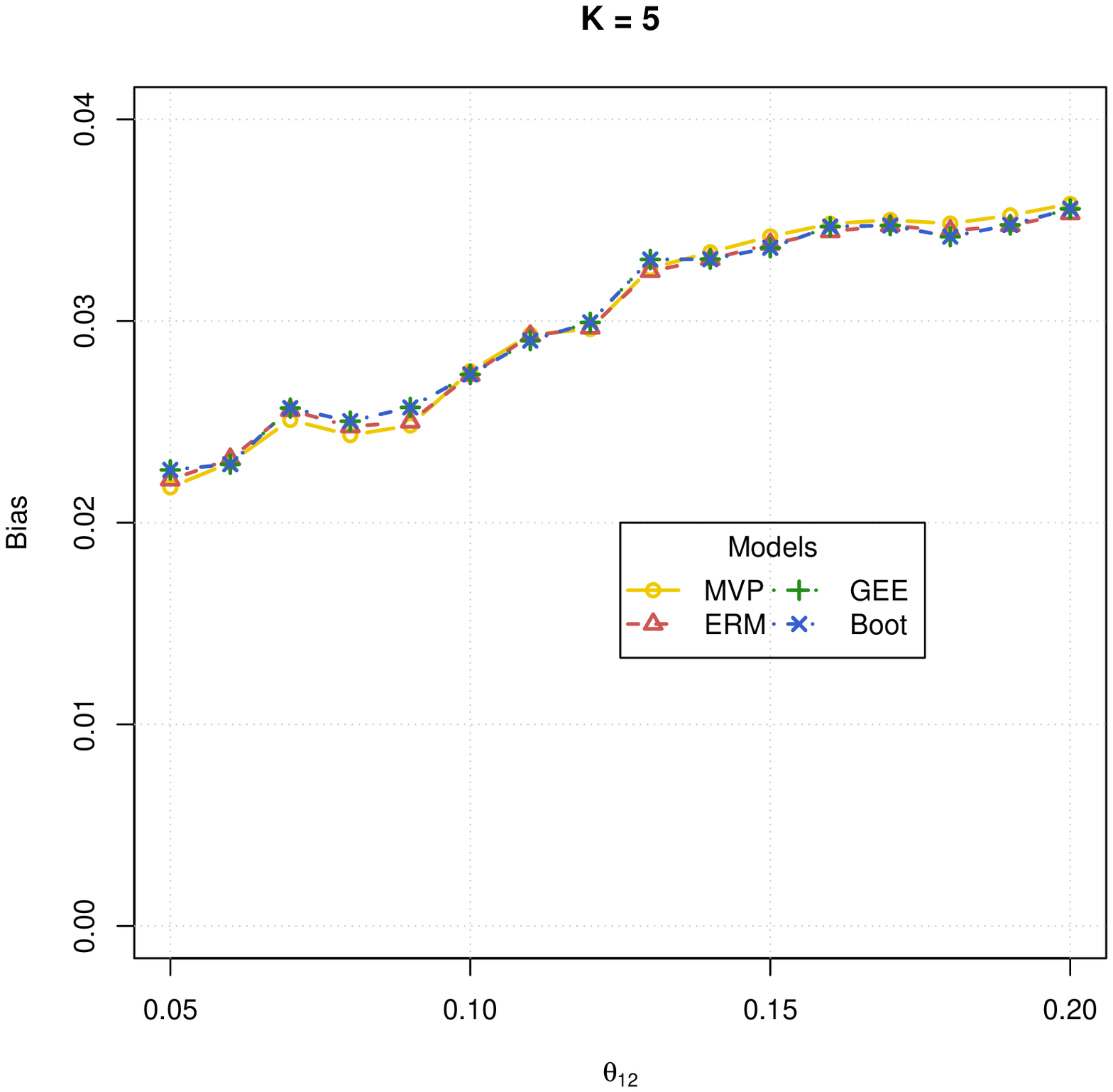}
\caption{Bias in estimating $\rho_2$ for the multivariate probit analysis (MVP), exponential risk model (ERM), GEE-based model (GEE), and the Bootstrap-based model (Boot) by $\theta_{12}$ and $K$.}\label{f:bias}
\end{figure}

\begin{figure}[h]%
\centering
\includegraphics[width=0.496\textwidth]{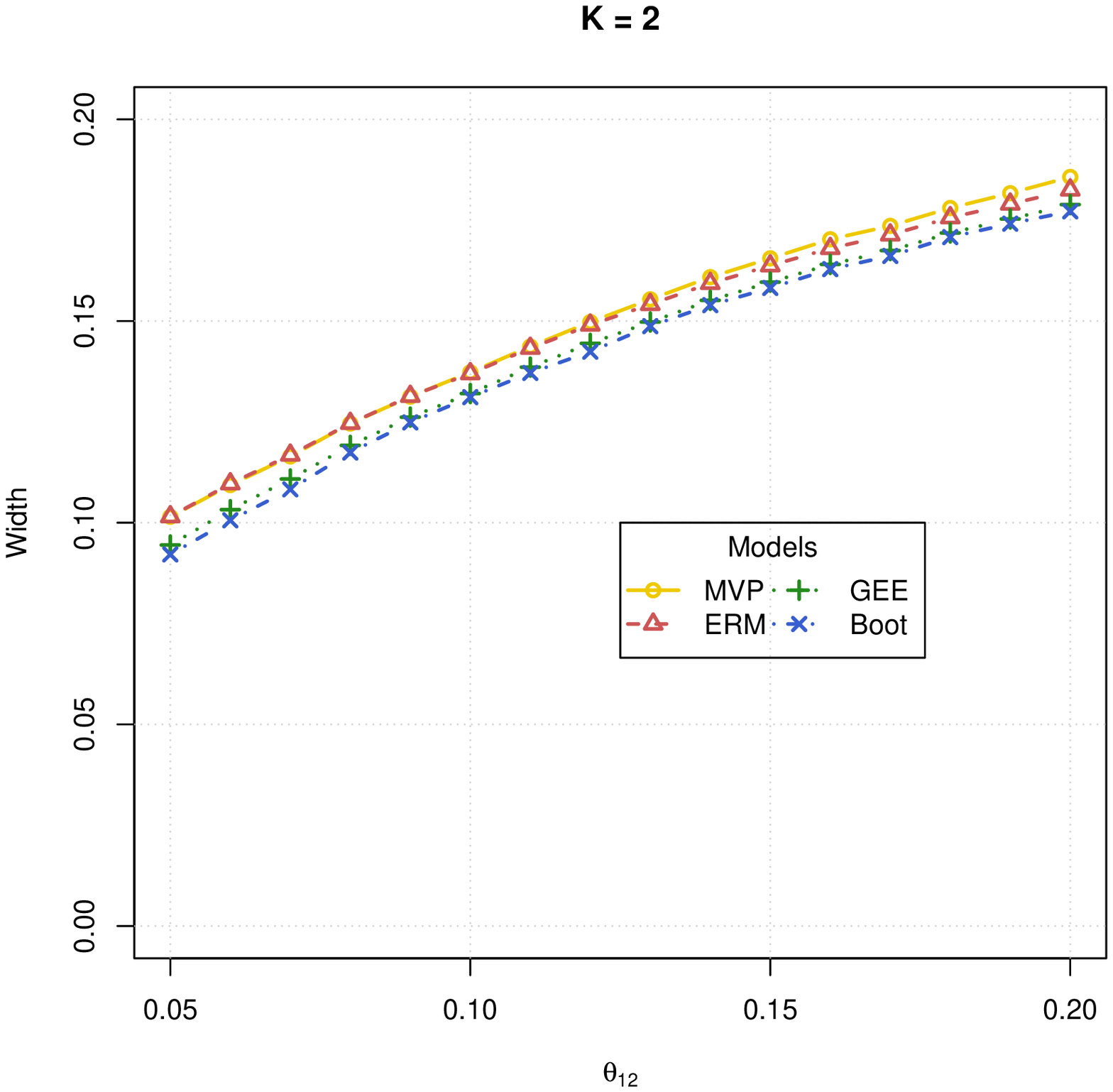}
\includegraphics[width=0.496\textwidth]{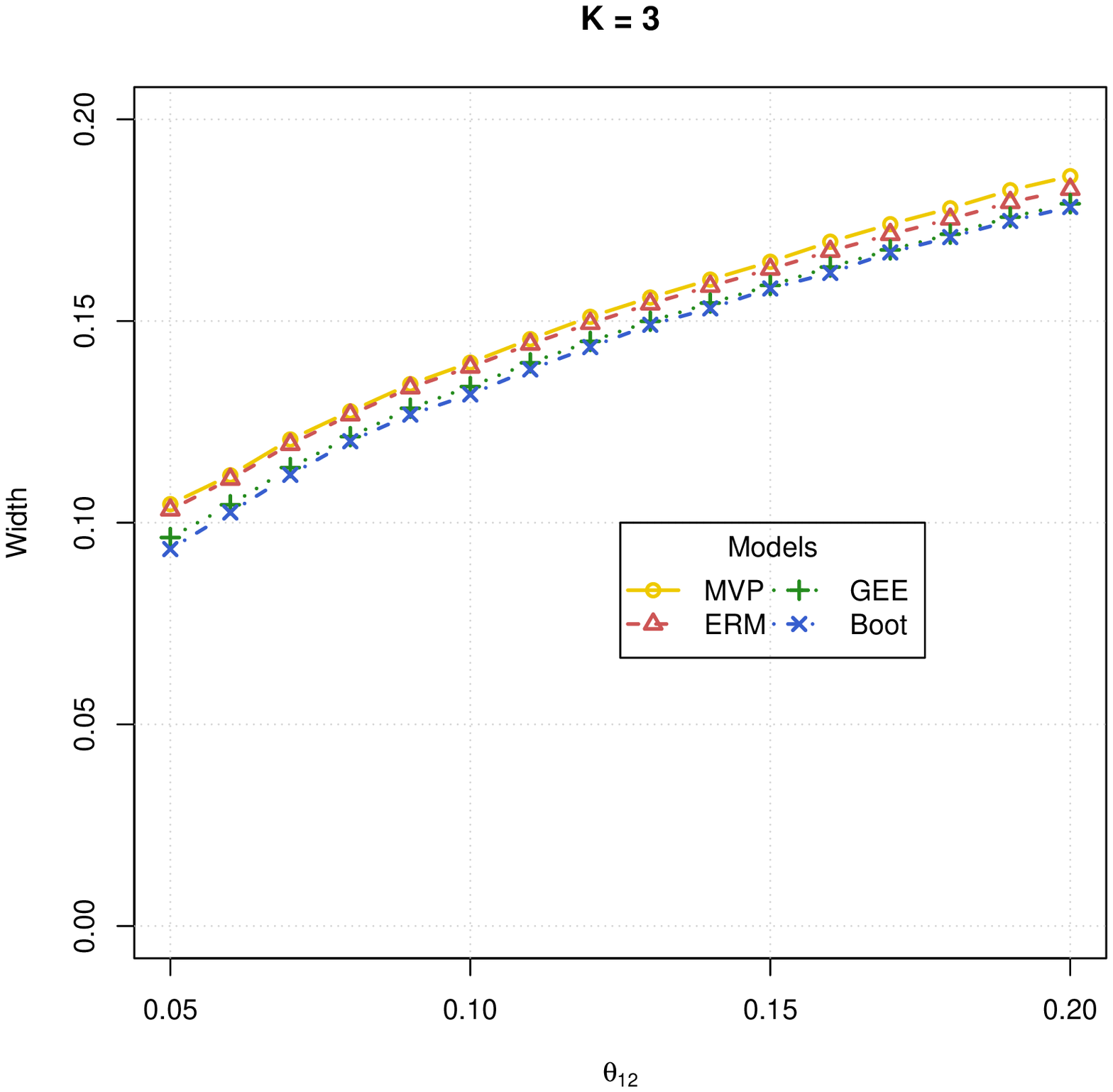}
\includegraphics[width=0.496\textwidth]{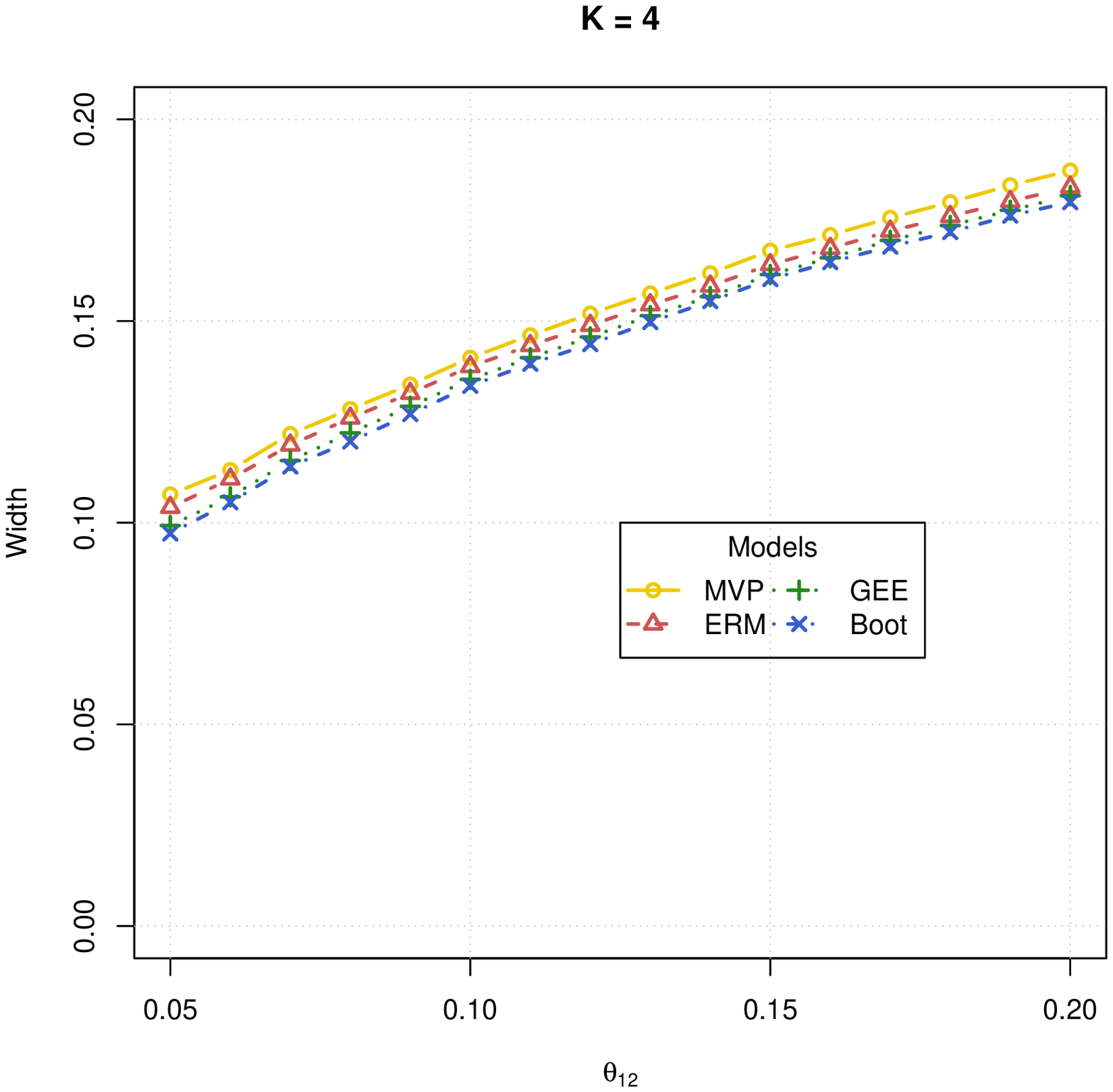}
\includegraphics[width=0.496\textwidth]{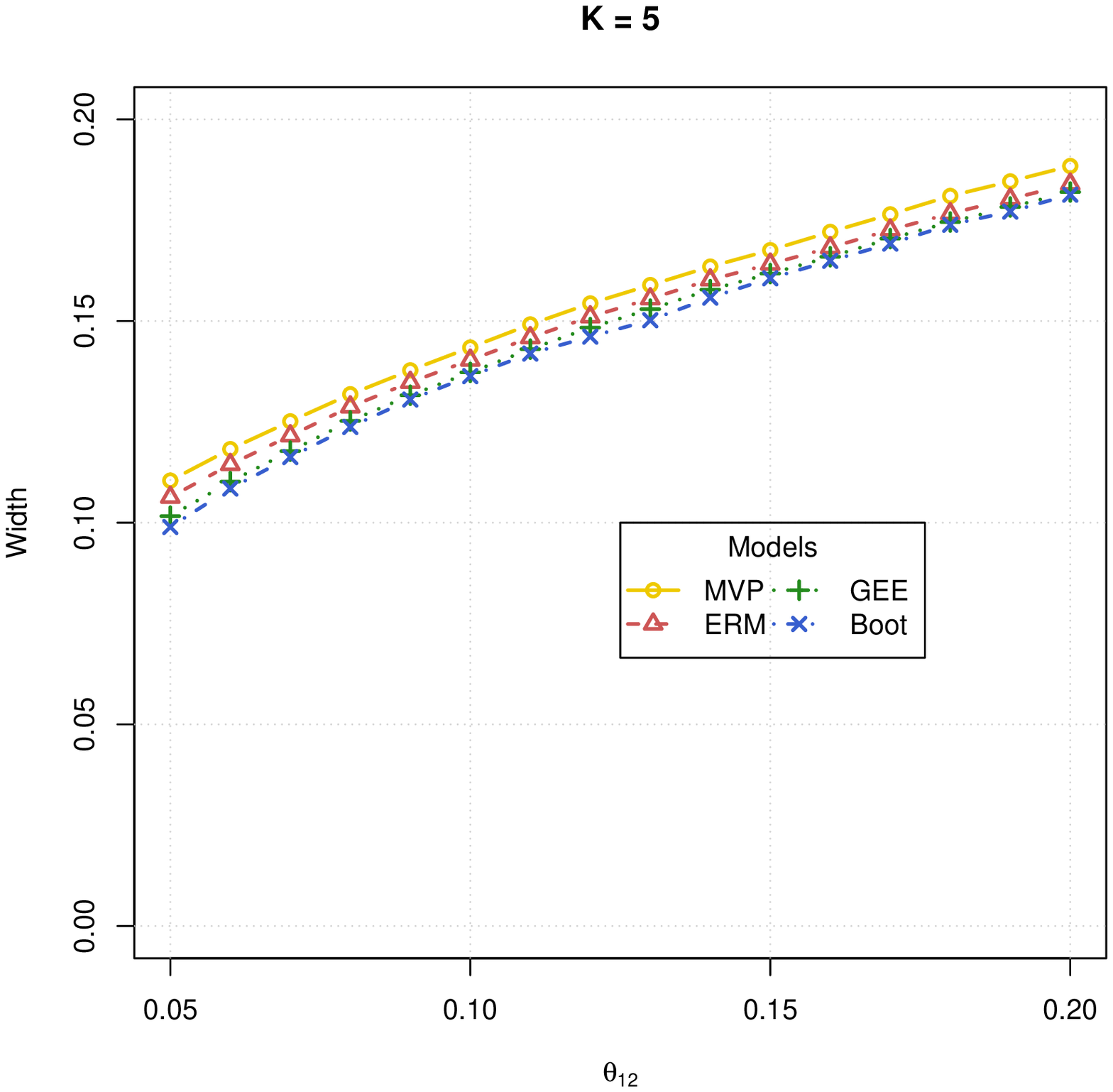}
\caption{Width of intervals on $\rho_2$ for the multivariate probit analysis (MVP), exponential risk model (ERM), GEE-based model (GEE), and the Bootstrap-based model (Boot) by $\theta_{12}$ and $K$. All intervals are 95\% intervals.}\label{f:width}
\end{figure}

Finally, we assess the bias of each method, in Figure~\ref{f:bias}, and interval width, in Figure~\ref{f:width}. The same color, line, and point schemes used in the coverage and power curves is used for the bias and width curves Bias is virtually identically between the four methods regardless of the value of $\theta_{12}$ or $K$. In general, bias tends to increase as $\theta_{12}$ and increases slightly with $K$ as well, though more so with changes in the effect size. Interval width is very similar between the four methods, although the multivariate analysis tends to have the widest intervals followed by the ERM, the GEE-based model, and then the Bootstrap-based method. For all methods, width increases as $\theta_{12}$ increases but only exhibits minimal changes as $K$ increases. Histograms of bias collapsing across $\theta_{12}$ are in Section 2 of the Supplementary Materials. 

\section{Re-analysis of the SOC Data}
\label{s:soc}

The goal of the SOC data is to determine if primary care physicians are initiating contact with the same agencies within the SOC framework, and with similar frequency, to psychiatric, or specialty, care. Knutson et al. argues that it is important to identify services for improved coordination to help the overburdened mental health system and their original findings suggest there is room for improvement among primary care givers \cite{Knutson2018}. They separately analyze only the non-sparse sets using standard univariate methods including logistic regression \cite{Knutson2018}. Because of the sparsity in the juvenile justice and developmental disabilities components, the authors do not examine these outcomes. Of the three outcomes they analyze, only contact with the education system, is deemed significantly different \cite{Knutson2018}. We now re-analyze all outcomes, simultaneously, using the multivariate analysis.

In total, we have 74 patients available for analysis. The most common primary mental health diagnosis was mood disorder followed by attention deficit hyperactivity disorder, adjustment disorder, and anxiety disorder \cite{Knutson2018}. Additional primary diagnoses include post-traumatic stress disorder, substance abuse, and learning disorder \cite{Knutson2018}. Each patient was initially seen by primary care, who made recommendations for contact (or not) with components of the SOC. Separately, specialty care later examined each patient and independently made recommendations for contact within the SOC. We only seek to estimate and analyze $\brho$, the differences in the marginal probability of contact between primary care and specialty care for each SOC component. Table~\ref{t:analysis} presents the results of the multivariate analysis including posterior median values for each $\rho_k$, 95\% credible intervals (CrI), and posterior probabilities that $\rho_k$ is greater than zero, $P(\rho_k > 0)$. In Table~\ref{t:analysis}, we abbreviate each component as follows: developmental disabilities (DD), mental health (MH), juvenile justice (JJ), child welfare (CW), and education system (ED). In total, we generated 20,000 draws from the posterior, retaining the last 10,000 for estimation and inference. All model parameters were judged to have converged using trace plots (see Section 3 of the Supplementary Materials) and the Gelman-Rubin diagnostic \cite{GelmanRubin1992} (see Table~\ref{t:analysis} where values of $\hat{R}$ and its upper 95\% CrI near one suggest convergence).

We begin by noting what is consistent in our analysis with the univariate analysis from Knutson et al.\cite{Knutson2018}. The authors found a significant association for the education component of the SOC whereby primary care contacted the education system less than specialty care. We confirm this result noting that the difference in the marginal probability of contact with the education system between primary care and specialty care is $\rho_{ED} = -0.394$ (95\% CrI $[-0.533, -0.240]$, $P[\rho_{ED} > 0] = 0.000$). Knutson et al.\cite{Knutson2018} also found no significant associations for the child welfare and mental health components. Our analysis confirms this result for contact with child welfare ($\rho_{CW} = -0.040$, 95\% CrI $[-0.150, 0.063]$,  $P[\rho_{CW} > 0] = 0.224$) but by analyzing the data using a multivariate analysis, we have a more nuanced view of the relationship between care type and contact with mental health. Specifically, the difference in contact with mental health between primary care and specialty care is $\rho_{MH} = -0.110$ and its 95\% CrI includes zero $(-0.259, 0.035)$. But only 6.8\% of the posterior distribution of $\rho_{MH}$ sits above zero which indicates that a large portion of the posterior distribution is negative.

\begin{table}[h]
	\caption{Results from the re-analysis of the SOC Data using the multivariate analysis. Gelman-Rubin $\hat{R}$ values and upper 95\% credible bounds near one suggest convergence. $P(\rho > 0)$ is the posterior probability that $\rho$ is greater than zero. Component abbreviations: developmental disabilities (DD), mental health (MH), juvenile justice (JJ), child welfare (CW), and education system (ED). \label{t:analysis}}
	\begin{center}
		\begin{tabular}{lcccccc}
			\toprule
			 &  &  \multicolumn{2}{c}{Cred. Int.} &  & \multicolumn{2}{c}{Gelman-Rubin} \\
			 \cmidrule{3-4} \cmidrule{6-7}
			Component & $\rho$  & 2.5\% & 97.5\% & $P(\rho > 0)$ & $\hat{R}$ & Upper 95\% \\
			\midrule
			DD &  0.035 & $-$0.002 & 0.096  &  0.970  & 1.003  &  1.008 \\
			MH & $-$0.110 & $-$0.259 & 0.035 &  0.068  & 1.000  &  1.001 \\
			JJ & $-$0.024 & $-$0.089 & 0.031  &  0.170  & 1.002  &  1.006 \\
			CW & $-$0.040 & $-$0.150 & 0.063  &  0.224  & 1.001  &  1.002\\
			ED & $-$0.394 & $-$0.533 & $-$0.240  &  0.000  & 1.000  &  1.001 \\
			\bottomrule
  		\end{tabular}
	\end{center}
\end{table}

Due to sparse response, contact with the developmental disabilities and juvenile justice components was not assessed in the original univariate analysis \cite{Knutson2018}. But using the proposed multivariate analysis, we can conduct inference on both. We do not find a significant association in contact with the juvenile justice component ($\rho_{JJ} = -0.024$, 95\% CrI $[-0.089, 0.031]$,  $P[\rho_{JJ} > 0] = 0.170$). The relationship between care type and contact with the developmental disabilities component is more nuanced. We find a positive effect suggesting that the marginal probability of contact is $\rho_{DD} = 0.035$ more for primary care than for specialty care. The 95\% CrI does exclude zero, $-0.002, 0.096$, although it is on the edge and the posterior probability that $\rho_{DD}$ is greater than zero is 97.0\%. Thus, a sizable portion of the posterior distribution of $\rho_{DD}$ is positive. Both the na\"ive and penalized analyses provide similar inferential results to the multivariate analysis (see Section 3 of the Supplementary Materials).

Overall, our analysis suggests there are inefficiencies in the system of care framework, a consistent finding with that of Knutson et al. \cite{Knutson2018}. But our results are more nuanced and suggest multiple areas for targeted improvement. In the study, primary care is recommending contact with both the education and mental health systems at lower rates than they should be; potentially leaving important actors out of the care for children with mental health disorders. Conversely, primary care is contacting disability services more than specialty care; potentially overburdening those services with children who do not need them. Many children only receive mental health care through their primary care physicians,\cite{Costello2014} so it is important these rates of contact to align to ensure the children benefit from the SOC framework\cite{Knutson2018}.

\section{Discussion}

Methods for MMP data are limited to non-Bayesian approaches that do not perform well in the presence of sparse response. In this manuscript, we propose several Bayesian analyses for MMP data including a multivariate analysis that models the latent covariance using a Bayesian FPCA. The Bayesian FPCA gives a low-dimensional expansion using well-behaved basis functions, in this case B-splines, that do not overly burden the multivariate analysis with parameters, whereas simulating from a general latent correlation would add a potentially large number of parameters to the model. Prior work demonstrates the ability of the Bayesian FPCA to adequately model a variety of complex covariance structures \cite{Goldsmith2016, MeyerMorris2022}. These structures are not unique to functional data and the technique can be used for multivariate data in general. The Bayesian FPCA is also useful for modeling latent covariances in multivariate probit models since those structures may be difficult to observe and therefore hard to pre-simplify with an assumption, e.g. a compound symmetric covariance instead of an unstructured one. Additionally, since the multivariate analysis is much more computationally efficient than other multivariate probit models, it is a useful tool for analyzing MMP data.

As we demonstrate in simulation, the multivariate analysis outperforms the existing methods in the presence of sparse response regardless of the effect size, as controlled by $\theta_{12}$, or the number of outcomes, $K$. It routinely has coverage in excess of nominal and does not dip below 92\% when constructing 95\% intervals. Conversely, the existing methods for MMP data have inconsistent coverage that in some cases can be as low as 84\% (again when constructing 95\% intervals). At the same time, the multivariate analysis has the highest power without sacrificing bias or coverage and with only minimally larger interval widths. Consequently, we recommend the use of the multivariate analysis for MMP data with sparse response. This ensures that the multivariate structure of the data is modeled, but we note that both the na\"ive and penalized models perform similarly to the multivariate analysis. All Bayesian methods we consider out perform the existing methods in the presence of sparse responses.

The re-analysis of the SOC data confirms earlier findings, specifically that contact with the education system is lower among primary care than specialty care. But our multivariate analysis reveals several additional results that warrant further investigation. Among the components that were previously analyzed, contact with the mental health component may also be lower for primary care than for specialty care. The main advantage our multivariate analysis provides is in the ability to also assess the sparse components which were not examined in the original univariate analysis. What we see is that there is some evidence of increased contact with disability services among primary care compared to specialty care. Taken all together, the SOC study and our re-analysis should provide the foundation for further examination of the coherence of SOC frameworks. This data can be thought of as a pilot study that can provide a good source for developing future, larger studies to further examine the differences in contact within the SOC between primary care and specialty care. Our Bayesian multivariate analysis can seamlessly incorporate prior information from this study by replacing the mean of the prior on $\bbeta$ with values derived from the analysis presented in Table~\ref{t:analysis}.

\backmatter

\bmhead{Supplementary Information}

The Supplementary Materials referenced in Sections~\ref{s:sim} and~\ref{s:soc} can be found alongside this manuscript.

\bmhead{Acknowledgments}

Partial funding for this work was provided by internal Georgetown University Summer Academic Research Grants.

\bmhead{Compliance with Ethical Standards}

The authors have no potential or actualy conflicts of interest to declare. No additional data was collected for this research. The original study was retrospective in nature and was approved by the Boston University Institutional Review Board \cite{Knutson2018}.

\bibliography{fullbib.bib}


\begin{thebibliography}{33}
\ifx \bisbn   \undefined \def \bisbn  #1{ISBN #1}\fi
\ifx \binits  \undefined \def \binits#1{#1}\fi
\ifx \bauthor  \undefined \def \bauthor#1{#1}\fi
\ifx \batitle  \undefined \def \batitle#1{#1}\fi
\ifx \bjtitle  \undefined \def \bjtitle#1{#1}\fi
\ifx \bvolume  \undefined \def \bvolume#1{\textbf{#1}}\fi
\ifx \byear  \undefined \def \byear#1{#1}\fi
\ifx \bissue  \undefined \def \bissue#1{#1}\fi
\ifx \bfpage  \undefined \def \bfpage#1{#1}\fi
\ifx \blpage  \undefined \def \blpage #1{#1}\fi
\ifx \burl  \undefined \def \burl#1{\textsf{#1}}\fi
\ifx \doiurl  \undefined \def \doiurl#1{\url{https://doi.org/#1}}\fi
\ifx \betal  \undefined \def \betal{\textit{et al.}}\fi
\ifx \binstitute  \undefined \def \binstitute#1{#1}\fi
\ifx \binstitutionaled  \undefined \def \binstitutionaled#1{#1}\fi
\ifx \bctitle  \undefined \def \bctitle#1{#1}\fi
\ifx \beditor  \undefined \def \beditor#1{#1}\fi
\ifx \bpublisher  \undefined \def \bpublisher#1{#1}\fi
\ifx \bbtitle  \undefined \def \bbtitle#1{#1}\fi
\ifx \bedition  \undefined \def \bedition#1{#1}\fi
\ifx \bseriesno  \undefined \def \bseriesno#1{#1}\fi
\ifx \blocation  \undefined \def \blocation#1{#1}\fi
\ifx \bsertitle  \undefined \def \bsertitle#1{#1}\fi
\ifx \bsnm \undefined \def \bsnm#1{#1}\fi
\ifx \bsuffix \undefined \def \bsuffix#1{#1}\fi
\ifx \bparticle \undefined \def \bparticle#1{#1}\fi
\ifx \barticle \undefined \def \barticle#1{#1}\fi
\bibcommenthead
\ifx \bconfdate \undefined \def \bconfdate #1{#1}\fi
\ifx \botherref \undefined \def \botherref #1{#1}\fi
\ifx \url \undefined \def \url#1{\textsf{#1}}\fi
\ifx \bchapter \undefined \def \bchapter#1{#1}\fi
\ifx \bbook \undefined \def \bbook#1{#1}\fi
\ifx \bcomment \undefined \def \bcomment#1{#1}\fi
\ifx \oauthor \undefined \def \oauthor#1{#1}\fi
\ifx \citeauthoryear \undefined \def \citeauthoryear#1{#1}\fi
\ifx \endbibitem  \undefined \def \endbibitem {}\fi
\ifx \bconflocation  \undefined \def \bconflocation#1{#1}\fi
\ifx \arxivurl  \undefined \def \arxivurl#1{\textsf{#1}}\fi
\csname PreBibitemsHook\endcsname

\bibitem{Knutson2018}
\begin{barticle}
\bauthor{\bsnm{Knutson}, \binits{K.H.}},
\bauthor{\bsnm{Meyer}, \binits{M.J.}},
\bauthor{\bsnm{Thakrar}, \binits{N.}},
\bauthor{\bsnm{Stein}, \binits{B.D.}}:
\batitle{Care coordination for youth with mental health disorders in primary
  care}.
\bjtitle{Clinical Pediatrics}
\bvolume{57},
\bfpage{5}--\blpage{10}
(\byear{2018}).
\doiurl{10.1177/0009922817733740}
\end{barticle}
\endbibitem

\bibitem{Kling2006}
\begin{barticle}
\bauthor{\bsnm{Klingenberg}, \binits{B.}},
\bauthor{\bsnm{Agresti}, \binits{A.}}:
\batitle{Multivariate extension of {McNemar's} test}.
\bjtitle{Biometrics}
\bvolume{62},
\bfpage{921}--\blpage{928}
(\byear{2006}).
\doiurl{10.1111/j.1541-0420.2006.00525.x}
\end{barticle}
\endbibitem

\bibitem{McNemar1947}
\begin{barticle}
\bauthor{\bsnm{McNemar}, \binits{Q.}}:
\batitle{Note on the sampling error of the difference between correlated
  proportions or percentages}.
\bjtitle{Pyschometrika}
\bvolume{12},
\bfpage{153}--\blpage{157}
(\byear{1947}).
\doiurl{10.1007/BF02295996}
\end{barticle}
\endbibitem

\bibitem{Consonni2008}
\begin{barticle}
\bauthor{\bsnm{Consonni}, \binits{G.}},
\bauthor{\bsnm{{La Rocca}}, \binits{L.}}:
\batitle{Tests based on intrinsic priors for the equality of two correlated
  proportions}.
\bjtitle{Journal of the American Statistical Association}
\bvolume{103},
\bfpage{1260}--\blpage{1269}
(\byear{2008}).
\doiurl{10.1198/01621450800000043}
\end{barticle}
\endbibitem

\bibitem{Saeki2017}
\begin{barticle}
\bauthor{\bsnm{Saeki}, \binits{H.}},
\bauthor{\bsnm{Tango}, \binits{T.}},
\bauthor{\bsnm{Wang}, \binits{J.}}:
\batitle{Statistical inference for noninferiority of difference in proportions
  of clustered matched-pair data from multiple raters}.
\bjtitle{Journal of Biopharmaceutical Statistics}
\bvolume{27},
\bfpage{70}--\blpage{83}
(\byear{2017}).
\doiurl{10.1080/10543406.2016.1148709}
\end{barticle}
\endbibitem

\bibitem{Westfall2010}
\begin{barticle}
\bauthor{\bsnm{Westfall}, \binits{P.H.}},
\bauthor{\bsnm{Troendle}, \binits{J.F.}},
\bauthor{\bsnm{Pennello}, \binits{G.}}:
\batitle{Multiple {McNemar} tests}.
\bjtitle{Biometrics}
\bvolume{66},
\bfpage{1185}--\blpage{1191}
(\byear{2010}).
\doiurl{10.1111/j.1541-0420.2010.01408.x}
\end{barticle}
\endbibitem

\bibitem{Xu2013}
\begin{barticle}
\bauthor{\bsnm{Xu}, \binits{J.}},
\bauthor{\bsnm{Yu}, \binits{M.}}:
\batitle{Sample size determination and re-estimation for matched pair designs
  with multiple binary endpoints}.
\bjtitle{Biometrical Journal}
\bvolume{55},
\bfpage{430}--\blpage{443}
(\byear{2013}).
\doiurl{10.1002/bimj.201100231}
\end{barticle}
\endbibitem

\bibitem{Lui2013}
\begin{barticle}
\bauthor{\bsnm{Lui}, \binits{K.-J.}},
\bauthor{\bsnm{Chang}, \binits{K.-C.}}:
\batitle{Testing and estimation of proportion (or risk) ratio under the
  matched-pair design with multiple binary endpoints}.
\bjtitle{Biometrical Journal}
\bvolume{55},
\bfpage{603}--\blpage{616}
(\byear{2013}).
\doiurl{10.1002/bimj.201200224}
\end{barticle}
\endbibitem

\bibitem{Lui2016}
\begin{barticle}
\bauthor{\bsnm{Lui}, \binits{K.-J.}},
\bauthor{\bsnm{Chang}, \binits{K.-C.}}:
\batitle{Notes on testing noninferiority in multivariate binary data under the
  matched-pair design}.
\bjtitle{Statistical Methods in Medical Research}
\bvolume{25},
\bfpage{1272}--\blpage{1289}
(\byear{2016}).
\doiurl{10.1177/0962280213477022}
\end{barticle}
\endbibitem

\bibitem{Cochran1950}
\begin{barticle}
\bauthor{\bsnm{Cochran}, \binits{W.G.}}:
\batitle{The comparison of percentages in matched samples}.
\bjtitle{Biometrika}
\bvolume{37},
\bfpage{256}--\blpage{266}
(\byear{1950}).
\doiurl{10.2307/2332378}
\end{barticle}
\endbibitem

\bibitem{Mantel1959}
\begin{barticle}
\bauthor{\bsnm{Mantel}, \binits{N.}},
\bauthor{\bsnm{Haenszel}, \binits{W.}}:
\batitle{Statistical aspects of the analysis of data from retrospective studies
  of disease}.
\bjtitle{Journal of the National Cancer Institute}
\bvolume{22},
\bfpage{719}--\blpage{748}
(\byear{1959}).
\doiurl{10.1093/jnci/22.4.719}
\end{barticle}
\endbibitem

\bibitem{Jiang2017}
\begin{barticle}
\bauthor{\bsnm{Jiang}, \binits{Y.}},
\bauthor{\bsnm{Xu}, \binits{J.}}:
\batitle{A comparative study of matched pair designs with two binary
  endpoints}.
\bjtitle{Statistical Methods in Medical Research}
\bvolume{26},
\bfpage{2526}--\blpage{2542}
(\byear{2017}).
\doiurl{10.1177/0962280215601136}
\end{barticle}
\endbibitem

\bibitem{Agresti2013}
\begin{bbook}
\bauthor{\bsnm{Agresti}, \binits{A.}}:
\bbtitle{Categorical Data Analysis},
\bedition{3$^{\text{rd}}$} edn.
\bpublisher{John Wiley \& Sons},
\blocation{Hoboken, NJ}
(\byear{2013})
\end{bbook}
\endbibitem

\bibitem{Altham1971}
\begin{barticle}
\bauthor{\bsnm{Altham}, \binits{P.M.E.}}:
\batitle{The analysis of matched proportions}.
\bjtitle{Biometrika}
\bvolume{58},
\bfpage{561}--\blpage{576}
(\byear{1971}).
\doiurl{10.2307/2334391}
\end{barticle}
\endbibitem

\bibitem{Broemeling1996}
\begin{barticle}
\bauthor{\bsnm{Broemeling}, \binits{L.D.}},
\bauthor{\bsnm{Gregurich}, \binits{M.A.}}:
\batitle{A {Bayesian} alternative to the analysis of matched categorical
  responses}.
\bjtitle{Communications in Statistics - Theory and Methods}
\bvolume{25},
\bfpage{1429}--\blpage{1445}
(\byear{1996}).
\doiurl{10.1080/03610929608831777}
\end{barticle}
\endbibitem

\bibitem{Ghosh2000}
\begin{barticle}
\bauthor{\bsnm{Ghosh}, \binits{M.}},
\bauthor{\bsnm{Chen}, \binits{M.-H.}},
\bauthor{\bsnm{Ghosh}, \binits{A.}},
\bauthor{\bsnm{Agresti}, \binits{A.}}:
\batitle{Hierarchical {Bayesian} analysis of binary matched pairs data}.
\bjtitle{Statistica Sinica}
\bvolume{10},
\bfpage{647}--\blpage{657}
(\byear{2000})
\end{barticle}
\endbibitem

\bibitem{AlbertChib1993}
\begin{barticle}
\bauthor{\bsnm{Albert}, \binits{J.}},
\bauthor{\bsnm{Chib}, \binits{S.}}:
\batitle{Bayesian analysis of binary and polychotomous response data}.
\bjtitle{Journal of the American Statistical Association}
\bvolume{88},
\bfpage{669}--\blpage{679}
(\byear{1993}).
\doiurl{10.1080/01621459.1993.10476321}
\end{barticle}
\endbibitem

\bibitem{AlbertChib1995}
\begin{barticle}
\bauthor{\bsnm{Albert}, \binits{J.}},
\bauthor{\bsnm{Chib}, \binits{S.}}:
\batitle{Bayesian residual analysis for binary response models}.
\bjtitle{Biometrika}
\bvolume{82},
\bfpage{747}--\blpage{759}
(\byear{1995}).
\doiurl{10.1093/biomet/82.4.747}
\end{barticle}
\endbibitem

\bibitem{Gelman2006}
\begin{barticle}
\bauthor{\bsnm{Gelman}, \binits{A.}}:
\batitle{Prior distributions for variance parameters in hierarchical models}.
\bjtitle{Bayesian Analysis}
\bvolume{1},
\bfpage{513}--\blpage{533}
(\byear{2006}).
\doiurl{10.1214/06-BA117A}
\end{barticle}
\endbibitem

\bibitem{VanDerLinde2008}
\begin{barticle}
\bauthor{\bparticle{der} \bsnm{Linde}, \binits{A.V.}}:
\batitle{Variational {Bayesian} functional {PCA}}.
\bjtitle{Computational Statistics \& Data Analysis}
\bvolume{53},
\bfpage{517}--\blpage{533}
(\byear{2008}).
\doiurl{10.1016/j.csda.2008.09.015}
\end{barticle}
\endbibitem

\bibitem{Gelman2013}
\begin{bbook}
\bauthor{\bsnm{Gelman}, \binits{A.}},
\bauthor{\bsnm{Carlin}, \binits{J.B.}},
\bauthor{\bsnm{Stern}, \binits{H.S.}},
\bauthor{\bsnm{Dunson}, \binits{D.B.}},
\bauthor{\bsnm{Vehtari}, \binits{A.}},
\bauthor{\bsnm{Rubin}, \binits{D.B.}}:
\bbtitle{Bayesian Data Analysis},
\bedition{3$^{\text{rd}}$} edn.
\bpublisher{Chapman and Hall-CRC},
\blocation{Boca Raton, FL}
(\byear{2013})
\end{bbook}
\endbibitem

\bibitem{Polson2012}
\begin{barticle}
\bauthor{\bsnm{Polson}, \binits{N.G.}},
\bauthor{\bsnm{Scott}, \binits{J.G.}}:
\batitle{On the half-{Cauchy} prior for a global scale parameter}.
\bjtitle{Bayesian Analysis}
\bvolume{7},
\bfpage{887}--\blpage{902}
(\byear{2012}).
\doiurl{10.1214/12-BA730}
\end{barticle}
\endbibitem

\bibitem{Wand2011}
\begin{barticle}
\bauthor{\bsnm{Wand}, \binits{M.P.}},
\bauthor{\bsnm{Ormerod}, \binits{J.T.}},
\bauthor{\bsnm{Padoan}, \binits{S..A.}},
\bauthor{\bsnm{Fr\"uhwirth}, \binits{R.}}:
\batitle{Mean field variational bayes for elaborate distributions}.
\bjtitle{Bayesian Analysis}
\bvolume{6},
\bfpage{847}--\blpage{900}
(\byear{2011}).
\doiurl{10.1214/11-BA631}
\end{barticle}
\endbibitem

\bibitem{ChibGreenberg1998}
\begin{barticle}
\bauthor{\bsnm{Chib}, \binits{S.}},
\bauthor{\bsnm{Greenberg}, \binits{E.}}:
\batitle{Analysis of multivariate probit models}.
\bjtitle{Biometrika}
\bvolume{85},
\bfpage{347}--\blpage{361}
(\byear{1998}).
\doiurl{10.1093/biomet/85.2.347}
\end{barticle}
\endbibitem

\bibitem{Liu2001}
\begin{barticle}
\bauthor{\bsnm{Liu}, \binits{C.}}:
\batitle{Discussion}.
\bjtitle{Journal of Computational and Graphical Statistics}
\bvolume{10},
\bfpage{75}--\blpage{81}
(\byear{2001}).
\doiurl{10.1198/10618600152418746}
\end{barticle}
\endbibitem

\bibitem{Zhang2006}
\begin{barticle}
\bauthor{\bsnm{Zhang}, \binits{X.}},
\bauthor{\bsnm{Boscardin}, \binits{W.J.}},
\bauthor{\bsnm{Belin}, \binits{T.R.}}:
\batitle{Sampling correlation matrices in {Bayesian} models with correlated
  latent variables}.
\bjtitle{Journal of Computational and Graphical Statistics}
\bvolume{15},
\bfpage{880}--\blpage{896}
(\byear{2006}).
\doiurl{10.1198/106186006X160050}
\end{barticle}
\endbibitem

\bibitem{Webb2008}
\begin{barticle}
\bauthor{\bsnm{Webb}, \binits{E.L.}},
\bauthor{\bsnm{Forster}, \binits{J.J.}}:
\batitle{Bayesian model determination for multivariate ordinal and binary
  data}.
\bjtitle{Computational Statistics \& Data Analysis}
\bvolume{52},
\bfpage{2632}--\blpage{2649}
(\byear{2008}).
\doiurl{10.1016/j.csda.2007.09.008}
\end{barticle}
\endbibitem

\bibitem{Goldsmith2016}
\begin{barticle}
\bauthor{\bsnm{Goldsmith}, \binits{J.}},
\bauthor{\bsnm{Kitago}, \binits{T.}}:
\batitle{Assessing systematic effects of stroke on motor control using
  hierarchical function-on-scalar regression}.
\bjtitle{Journal of the Royal Statistical Society, Series C}
\bvolume{65},
\bfpage{215}--\blpage{236}
(\byear{2016}).
\doiurl{10.1111/rssc.12115}
\end{barticle}
\endbibitem

\bibitem{MeyerMorris2022}
\begin{barticle}
\bauthor{\bsnm{Meyer}, \binits{M.J.}},
\bauthor{\bsnm{Morris}, \binits{J.S.}},
\bauthor{\bsnm{Gazes}, \binits{R.P.}},
\bauthor{\bsnm{Coull}, \binits{B.A.}}:
\batitle{Ordinal probit functional outcome regression with application to
  computer-use behavior in rhesus monkeys}.
\bjtitle{Annals of Applied Statistics}
\bvolume{16},
\bfpage{537}--\blpage{550}
(\byear{2022}).
\doiurl{10.1214/21-AOAS1513}
\end{barticle}
\endbibitem

\bibitem{AgrestiCoull1998}
\begin{barticle}
\bauthor{\bsnm{Agresti}, \binits{A.}},
\bauthor{\bsnm{Coull}, \binits{B.A.}}:
\batitle{Approximate is better than 'exact' for interval estimation of binomial
  proportions}.
\bjtitle{The American Statistician}
\bvolume{52},
\bfpage{119}--\blpage{126}
(\byear{1998}).
\doiurl{10.1080/00031305.1998.10480550}
\end{barticle}
\endbibitem

\bibitem{AgrestiCaffo2000}
\begin{barticle}
\bauthor{\bsnm{Agresti}, \binits{A.}},
\bauthor{\bsnm{Caffo}, \binits{B.}}:
\batitle{Simple and effective confidence intervals for proportions and
  differences of proportions result from adding two successes and two
  failures}.
\bjtitle{The American Statistician}
\bvolume{54},
\bfpage{280}--\blpage{288}
(\byear{2000}).
\doiurl{10.1080/00031305.2000.10474560}
\end{barticle}
\endbibitem

\bibitem{GelmanRubin1992}
\begin{barticle}
\bauthor{\bsnm{Gelman}, \binits{A.}},
\bauthor{\bsnm{Rubin}, \binits{D.B.}}:
\batitle{Inference from iterative simulation using multiple sequences}.
\bjtitle{Statistical Science}
\bvolume{7},
\bfpage{457}--\blpage{511}
(\byear{1992}).
\doiurl{10.1214/ss/1177011136}
\end{barticle}
\endbibitem

\bibitem{Costello2014}
\begin{barticle}
\bauthor{\bsnm{Costello}, \binits{E.J.}},
\bauthor{\bsnm{He}, \binits{J.-P.}},
\bauthor{\bsnm{Sampson}, \binits{N.A.}},
\bauthor{\bsnm{Kessler}, \binits{R.C.}},
\bauthor{\bsnm{Merikangas}, \binits{K.R.}}:
\batitle{Services for adolescents with psychiatric disorders: 12-month data
  from the {National Comorbidity Survey-Adolescent}}.
\bjtitle{Psychiatric Services}
\bvolume{65},
\bfpage{359}--\blpage{366}
(\byear{2014}).
\doiurl{10.1176/appi.ps.201100518}
\end{barticle}
\endbibitem

\end{thebibliography}


\end{document}